\documentclass[12pt,preprint]{emulateapj}
\usepackage{amsmath}
\usepackage{graphicx}

\newcommand{\alphaox}{\alpha_{\rm OX}}
\newcommand{\fuv}{F_{\rm UV}}
\newcommand{\luv}{L_{\rm UV}}
\newcommand{\fx}{F_{\rm X}}
\newcommand{\lx}{L_{\rm X}}
\newcommand{\nh}{N_{\rm H}}
\newcommand{\fluxunits}{{\rm erg}\;{\rm s}^{-1}{\rm cm}^{-2}}

\newcommand{\fxsp}{F_{\rm [2-10]keV, sp}}
\newcommand{\rev}[1]{{#1}}

\shorttitle{A Hubble diagram for quasars}
\shortauthors{Risaliti \& Lusso}

\begin{document}


\title{A Hubble Diagram for Quasars}


\author{G. Risaliti\altaffilmark{1} and E. Lusso\altaffilmark{1}}
\affil{INAF - Arcetri Astrophysical Observatory, Largo E. Fermi 5, I-50125 Firenze, Italy}


\begin{abstract}
We present a new method to test the cosmological model, and to estimate the cosmological parameters, based on the non-linear relation between ultraviolet and X--ray luminosity of quasars. We built a data set of 1,138 quasars by merging several literature samples with X--ray measurements at 2~keV and SDSS photometry, which was used to estimate the extinction-corrected 2500~\AA\ flux. We obtained three main results: (1) we checked the non-linear relation between X--ray and UV luminosities in small redshift bins up to $z\sim6$, confirming that it holds at all redshifts with the same slope; (2) we built a Hubble diagram for quasars up to $z\sim6$, which is well matched to that of supernovae in the common $z=0-1.4$ redshift interval, and extends the test of the cosmological model up to $z\sim6$; (3) we showed that this non-linear relation is a powerful tool to estimate cosmological parameters. With present data, assuming a $\Lambda$CDM model, we obtain $\Omega_M$=0.22$^{+0.10}_{-0.08}$ and $\Omega_\Lambda$=0.92$^{+0.18}_{-0.30}$ ($\Omega_M$=0.28$\pm$0.04 and $\Omega_\Lambda$=0.73$\pm0.08$ from a joint quasar-SNe fit). Much more precise measurements will be achieved with future surveys. A few thousands SDSS quasars already have serendipitous X--ray observations with {\em Chandra} or XMM--{\em Newton}, and at least 100,000 quasars with UV and X--ray data will be available from the eROSITA all-sky survey in a few years. {\em Euclid}, LSST, and {\em Athena} surveys will further increase the sample size to at least several hundred thousands. Our simulations show that these samples will provide tight constraints on the cosmological parameters, and will allow to test possible deviations from the standard model with higher precisions than available today. 
\end{abstract}





\section{Introduction}

An optimal ``standard candle'' for cosmological studies is defined by two fundamental properties: it has a standard (or standard-izable) luminosity, and it is easy to observe in a wide redshift range. 
Quasars are the best class of astrophysical sources concerning the latter property, but completely lack the former: their observed emission spans several orders of magnitudes in luminosity, but their spectral energy distribution (SED) show no, or little, significant evolution with luminosity. As a consequence, quasars at first glance are far from being useful tools to set the cosmic distance scale as a function of redshift (i.e. the "Hubble Diagram"). 

However, even weak correlations between spectral features and luminosity, with large dispersions and observational biases, can in principle be useful for cosmological measurements, provided that the quasar sample is large enough. This consideration prompted several attempts to derive cosmological parameters from quasar observations. Examples are the anti-correlation between UV emission lines and luminosity \citep{baldwin1977,osmer-shields}, the luminosity-mass relation in super-Eddington accreting quasars \citep{wang2014}, the relation between luminosity and X--ray variability \citep{lafranca2014}, the radius-luminosity relationship \citep{watson2011,melia2014,kilerci2015}. 

In this paper we will explore the possibility of building a Hubble diagram for quasars using the well known non-linear relation between the UV and X--ray luminosities. 
\rev{ In recent years, several quasar samples have been observed at both optical/UV and X-ray wavelengths. We will show that if we collect all the available data in the literature, }
we can build a large enough sample to significantly constrain the cosmological parameters, and to test the cosmological model over the whole redshift range out to $z=0-6.6$ (age=0.8 Gyr). This analysis is achievable thanks to the increase of the number of quasars with observed UV and X--ray emission (provided by recent optical and X--ray surveys) by about an order of magnitude.

A non-linear relation between the UV and X--ray luminosity in quasars has been first discovered with the first X--ray surveys \citep{avnitananbaum79,zamorani81,avnitananbaum86}, and has been confirmed with various samples of a few hundred quasars observed with the main X--ray observatories over a redshift range from 0 to 6.5 and about five decades in UV luminosity. The largest samples recently analyzed include (1) a compilation of 367 quasars from different optical surveys and observed by {\em ROSAT, XMM-Newton} and {\em Chandra} (333 from \citealt{steffen06} and 34 from \citealt{just07}); (2) a sample of 350 sources obtained from the Sloan Digital Sky Survey second quasar catalog with {\em XMM-Newton} observations \citep{young10}, (3) a sample of 545 object from the COSMOS--XMM{\em-Newton} survey \citep{lusso2010}; (4) a sample of 200 quasars with UV and X--ray observations from the {\em Swift} observatory \citep{grupe10,vagnetti2010}.

In all these works the $\lx$--$\luv$ relation is parametrized as a linear dependence between the logarithm of the monochromatic luminosity at 2500~\AA\ ($\luv$) and the $\alphaox$ parameter, defined as the slope of a power law connecting the monochromatic luminosity at 2~keV ($\lx$), and L$_{UV}$: $\alphaox$=0.384$\times\log(\lx/\luv)$.  Luminosities are derived from fluxes through a luminosity distance calculated adopting a standard $\Lambda$CDM model with the best estimates of the cosmological parameters $\Omega_M$ and $\Omega_\Lambda$ at the time of the publications. When expressed as a relation between X--ray and UV luminosities, the $\alphaox$--$\luv$ relation becomes: $\log \lx=\beta+\gamma\log \luv$, with all the works cited above providing consistent values of the free parameters: $\beta\sim$9 and $\gamma\sim$0.6. The observed dispersion is $\delta\sim$0.35--0.40.

The potential use of this relation as a cosmological probe is obvious: if we assume no redshift evolution of the relation, the observed X--ray flux is a function of the observed UV flux, the redshift, and the parameters of the adopted cosmological model. The relation can be then fitted to a set of UV and X--ray observations of quasars in order to estimate the cosmological parameters.

However, none of the samples published so far has the size and/or homogeneity in the observations to provide useful constraints on cosmological parameters. \rev{ In this paper we bring together the largest quasar samples  available so far with both optical/UV and X-ray observations, in order to build a sample with broad luminosity and redshift coverage, and a large enough size to obtain a meaningful estimate of the cosmological parameters. }

This paper is organized as follows. 
In Section~\ref{Method} we will describe the method adopted to obtain the Hubble Diagram for quasars, and to estimate cosmological parameters (details are given in Appendix~\ref{appendixB}). In Section~\ref{The Sample} we discuss the general criteria for sample selection (with a complete discussion presented in Appendix~\ref{appendixA}). In Section~\ref{Analysis of the lum relation} we analyze the $\lx-\luv$ in our sample, and we validate its use as a cosmological probe. In Section~\ref{Determination of the cosmological parameters} we present the main results: the Hubble Diagram for quasars, and the measurements of cosmological parameter it provides. We also compare and complement our results with those from supernovae, in particular by comparing the Hubble diagram of supernovae and quasars below $z$$\sim$1.4, and calculating the cosmological parameters from joint fits. Our method and results are discussed in Section~\ref{Discussion}, while in Section~\ref{Future developments} we will explore possible future extensions of our work, based on the inclusion of more already available sources to our sample, on new dedicated observations, and on forthcoming new surveys. Conclusions are outlined in Section~\ref{Conclusions}. 

For luminosity estimates we adopted a concordance $\Lambda$-cosmology with $H_{0}=70\, \rm{km \,s^{-1}\, Mpc^{-1}}$, $\Omega_{M}=0.3$, $\Omega_{\Lambda}=1-\Omega_{M}$ \citep{komatsu09}.

\section{Method}
\label{Method}

Our method is based on the non-linear relation between $\lx$ and $\luv$:
\begin{equation}
\label{eq:luvlx}
\log(\lx)=\beta+\gamma\log(\luv).
\end{equation}
From the we above equation we obtain
\begin{equation}
\label{eq:fuvfx}
\log(\fx)=\Phi(F_{UV},D_L)=\beta'+\gamma\log(\fuv)+2(\gamma-1)\log(D_L),
\end{equation}
where $\beta'$ depends on the slope and intercept (i.e. $\beta'=\beta+(\gamma-1)\log(4\pi)$), $\fx$ and $\fuv$ are measured at fixed rest-frame wavelengths, and $D_{L}$ is the luminosity distance, which in a standard $\Lambda$CDM 
model (i.e. in a cosmological model with a fixed cosmological constant $\Lambda$) 
is given by
\begin{equation}
\begin{split}
\label{eq:dl}
D_L(z,\Omega_M,\Omega_\Lambda)=\frac{(1+z)}{\sqrt{\Omega_K}} \sinh \sqrt{\Omega_K}\\
\times\int^z_0\frac{dz}{H_0\sqrt{\Omega_M(1+z)^3+\Omega_\Lambda+\Omega_K(1+z)^2}},
\end{split}
\end{equation}
where $\Omega_K=1-\Omega_M-\Omega_\Lambda$.

Our analysis will consist of two main parts:\\
{\bf 1) Validation of the relation at different redshifts.} It is clear that any use of the above equations to determine the cosmological parameters is based on the assumption of no evolution of the $\lx$--$\luv$ relation. While it is not possible to check whether a redshift dependence of the scaling parameter $\beta$ is present (unless we have a physical model to determine it independently of the observed data, which is currently not the case, as discussed in Section~\ref{Discussion}), we can test the linear shape of the correlation and its slope $\gamma$ at different redshifts. 
In order to do so, we note that if Equation~\ref{eq:fuvfx} is analyzed in a sufficiently narrow redshift interval, the term containing $D_{L}$ will provide a nearly constant contribution. 
In particular, if its range of values within the chosen redshift interval is smaller than the intrinsic dispersion of the correlation, we will be able to test the relation by replacing luminosities with observed fluxes. 
An analysis of the function $D_{L}(z)$ for several values of $\Omega_{M}$ and $\Omega_\Lambda$ shows that in every redshift interval at $z>0.3$ with amplitude  $\Delta[\log z]$ we have $\Delta[\log D_L] < 0.7\Delta[\log z]$.

The measured dispersion of the global $\lx$--$\luv$ in the main published works cited in the Introduction is of the order of $\delta\sim$0.35--0.40. In order to have a negligible contribution by the distance term $D_L$ in a flux-flux relation, we need $\Delta[\log D_L] <0.10-0.15$. This implies a maximum size of the redshift bins $\Delta[\log z] < 0.1$. This sets a first important requirement for a quasar sample in order to be well suited for cosmological studies: it should contain enough sources to allow a significant test of the $\fx$--$\fuv$ relation in redshift bins smaller than $\Delta[\log z]\sim$0.1. \\
{\bf 2) Determination of the cosmological parameters}.
We can derive the cosmological parameters from the relation in Equation~\ref{eq:fuvfx} in two equivalent 
ways:
\begin{itemize}
\item We fit Equation~\ref{eq:fuvfx} by minimizing a likelihood function ($LF$) consisting of a modified $\chi^2$ function, allowing for an intrinsic dispersion $\delta$:
\begin{equation}
\label{eq:LF}
\ln(LF)=-\sum^N_{i=1}\left\{\frac{[\log(\fx)_i-\Phi(F_{UV},D_L)_i]^2}{s_i^2}
+\ln(s_i^2)\right\}
\end{equation}
where $\Phi(F_{UV},D_L)_i$ is given by Eq.~2, and $s_i^2=\sigma_i^2+\delta^2$, with $\sigma_i$ and $\delta$ indicating the measurement errors over $\fx$ and the global intrinsic dispersion, respectively. \rev{ We note that the dispersion} $\delta$ is much higher than typical values of $\sigma_i$, hence the fit is almost insensitive to the exact $\sigma_i$ value. The results do not change if an additional error due to systematic effects is added to $\sigma_i$, provided it does not exceed the intrinsic dispersion.
In this case the free parameters are $\delta$, the slope and intercept 
$\gamma$ and $\beta$, and the cosmological parameters $\Omega_{M}$ 
and $\Omega_\Lambda$ \rev{(we assume no evolution of the equation of state of dark energy, i.e. $w=-1$)}. We note that the Hubble constant $H_{0}$ 
is absorbed into the parameter $\beta$: without an independent determination of 
this parameter, our fits are insensitive to the value of $H_{0}$. 

\item We can use the best fit value and the uncertainty of the slope $\gamma$ (estimated 
in narrow redshift bins, as described above) to directly compute the distance modulus 
for each quasar in the sample, simply rearranging Equation~\ref{eq:fuvfx} in order to obtain the 
luminosity distance as a function of the observed X--ray and UV fluxes. We then 
fit $D_{L}(\Omega_M,\Omega_\Lambda)$ as a function of redshift, minimizing a likelihood analogous to the one in Equation~\ref{eq:LF}.
In this case the free parameters are the intrinsic dispersion, the cosmological parameters $\Omega_M$ and $\Omega_\Lambda$, and a residual scaling parameter $\beta'$.
\end{itemize}
The two methods outlined above are equivalent and provide fully consistent results. 
However, we note that a direct fit of the luminosity distance as a function of redshift has two practical advantages: its easier visualization, being formally a fit of a function of one variable, while the previous method requires a fit of $\fx$ as a function of ($\fuv,z$), and its homogeneity with other cosmological probes using Hubble Diagrams (chiefly those based on supernovae Ia).

The minimization has been performed using a Monte Carlo Markov Chain (MCMC), with the \textsc{emcee} package \citep{2013PASP..125..306F} in Python 2.7. Flat priors were used for each parameter. In order to further check the results, we also fitted the data in two other ways: (1) we used a standard likelihood maximization, based on the Levenberg-Marquardt (LM) algorithm; (2)  we performed a ``brute force'' maximization calculating the likelihood function in a grid of $\Omega_{M}$, $\Omega_\Lambda$, and $\delta$ (the $\gamma$ and $\beta$ values minimizing $LF$ can be computed analytically as a function of the other three parameters). In all cases, we find fully consistent results. In the latter approach, the confidence contours have been estimated via a bootstrap procedure. While the grid method is not efficient for precise parameter estimates (and has been used only to check the shape of the likelihood function), both the MCMC and LM methods have advantages (in particular, the latter is faster). Here we prefer to report the results from the MCMC approach for consistency with the main published works on observational cosmology.

\section{The sample}
\label{The Sample}
To obtain an adequate coverage of both the $\fuv-z$ and $\fx-z$ planes we need to combine wide, narrow, and deep-field surveys. In this first work we concentrated on data already available in the literature from published works on the $\alphaox-\luv$ correlation, with the requirement of a good photometric coverage in the optical-UV, in order to obtain precise estimates of F$_{UV}$ and of the possible extinction due to dust. We started with the quasar sample presented by \citet{steffen06}, which is a collection of several surveys such as the wide-field Sloan Digital Sky Survey (SDSS) quasar sample, low-redshift ($z < 0.2$) Seyfert 1 galaxies, and high-redshift ($z > 4$) optically selected AGNs.
We have expanded this sample with the quasars at $z>4$ by \citet{shemmer06}, the optically selected sample of high luminous quasars by \citet{just07} and \citet{young10}. These optically selected samples are then combined with the X--ray selected one analysed by \citet{lusso2010} for a total of $\sim$1,138 sources (see appendixes~\ref{A.1} and \ref{A.2} for further details).

All these catalogs provide a measurement of the rest-frame X--ray flux at 2~keV ($\fx$), and the UV flux at 2500\AA\ ($\fuv$). 
In our work, we adopted the published X--ray fluxes for most sources, while we updated the $\fuv$ estimates by using whenever possible the available multi-color information compiled in the COSMOS, SDSS-DR7, and BOSS-DR10 quasar catalogs. 
While we refer the reader to Appendix~\ref{appendixA} for details on our sample, flux estimates and possible systematics, a brief summary is given below. 

The adopted catalogs include multi-wavelength data from mid-infrared to ultraviolet: MIPS 24 $\mu$m GO3 data, IRAC flux densities, near-infrared Y-J-H-K-bands (2MASS and/or UKIDSS), optical photometry (e.g. SDSS, Subaru, CFHT), and near- and far-UV bands (GALEX). Observed magnitudes are converted into fluxes and corrected for Galactic reddening by employing a selective attenuation of the stellar continuum with $R_{V}=3.1$. 
Galactic extinction is estimated individually for each object in all catalogs.
For each source we considered the flux and corresponding effective frequency in 
each of the available bands. The data for the SED computation from mid-infrared 
to UV (upper limits are not considered) were then blueshifted to the rest-frame 
and no K-correction has been applied. We determine a ``first order" SED 
by using a first order polynomial function (i.e. straight line in the $\log \nu-\log (\nu F_{\nu})$ plane), which allows us to build densely sampled 
SEDs at all frequencies. This choice is motivated by the fact that a single interpolation 
with a high-order polynomial function could introduce spurious features in the 
final SED. $\fuv$ are extracted from the rest-frame SEDs in 
the $\log\nu-\log(\nu F_\nu)$ plane. In the case data do not cover 2500~\AA, fluxes 
are extrapolated at lower (higher) frequencies by considering, at least, the last 
(first) two photometric data points. 

Finally, we corrected the $\fuv$ estimates taking into account the redshift-dependent contribution of emission lines to the photometric points, as discussed in Appendix~\ref{A.4}.
An analogous correction to the values of $\fx$ based on the comparison between our values and the results of a complete X--ray spectral analysis is discussed in Appendix~\ref{A.5}.

\subsection{Source selection}
\label{Source selection}
\begin{figure}
 \resizebox{\hsize}{!}{\includegraphics{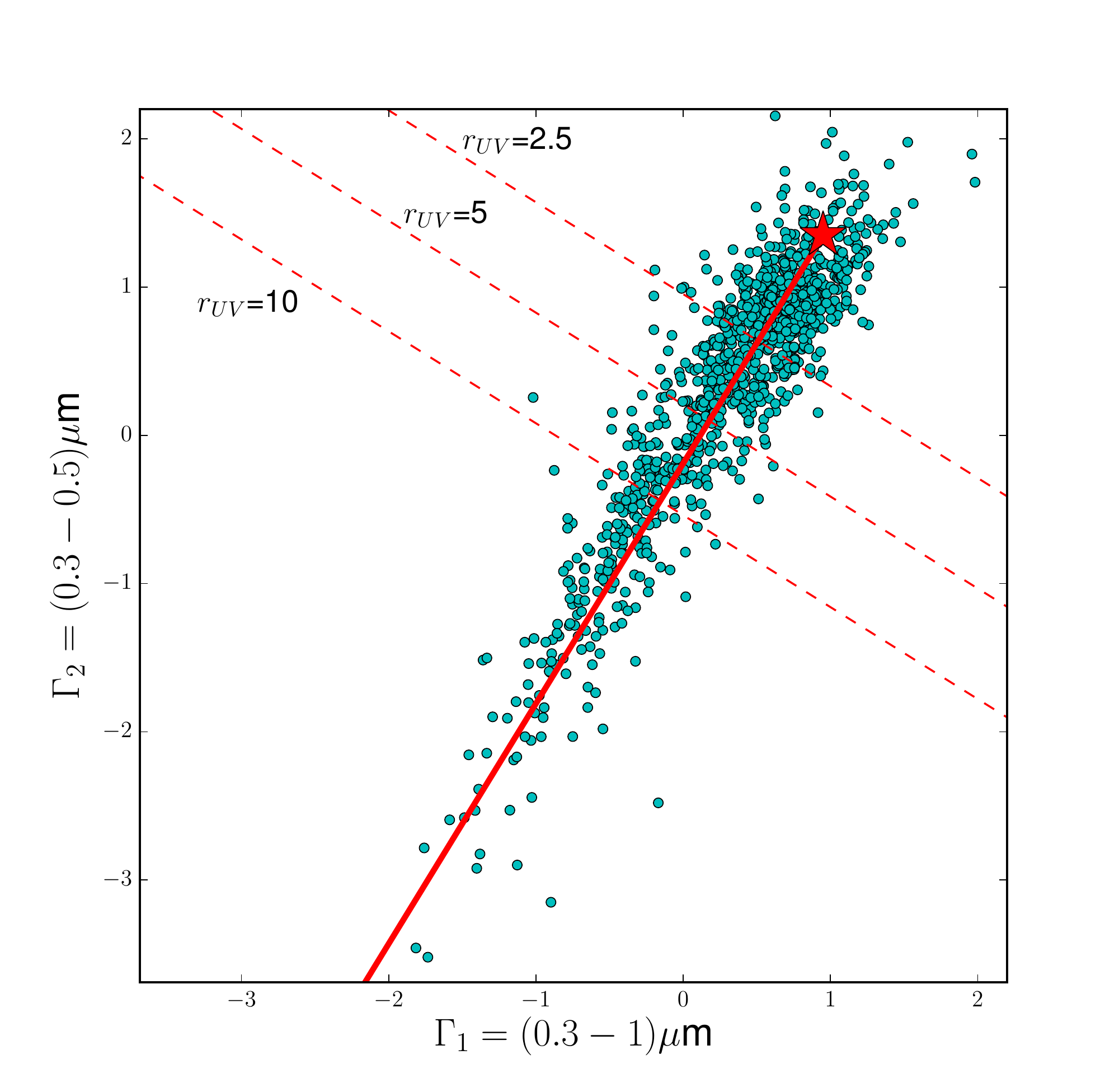}}
\caption{Distribution of the whole quasar sample (excluding radio-loud and BAL sources) in a $\Gamma_{1}-\Gamma_{2}$ plot, where $\Gamma_{1}$ and $\Gamma_{2}$ are the slopes of a power law in the $\log\nu-\log(\nu F_\nu)$ plane, in the 0.3--0.5~$\mu$ and 0.3--1~$\mu$m intervals, respectively. \rev{ The typical error for $\Gamma_1$ and $\Gamma_2$ is of the order of 0.1.} The red star represents the intrinsic quasar SED as estimated by \citet{2006AJ....131.2766R}. The solid red line is obtained by assuming increasing dust extinction following the extinction law of \citet{prevot84} and therefore, even if it nicely reproduces the correlation among the two colors, is {\it not} a fit to the points. The dashed lines, orthogonal to the solid one, show the colors corresponding to different values of the ratio $r_{UV}$  between the intrinsic and observed flux at 2500~\AA.  
\label{fig1}}
\end{figure}
The whole group of quasars collected as described above is not yet well suited for 
cosmological studies. 
The most obvious sources deviating from the $\lx$--$\luv$ relation are the heavily obscured Broad Absorption Line (BAL) and radio-loud quasars which have an additional, jet-linked, X-ray component. These sources were already removed by most of the parent samples, with a few exceptions (mostly radio-loud objects) which we further excluded from our sample.
A few undetected BAL quasars may still be present at low redshift. These objects may introduce a small bias in the $\lx$--$\luv$ relation\footnote{BALs are known to be X--ray obscured (e.g., \citealt{green1995ApJ...450...51G}, \citealt{1999ApJ...519..549G}, \citealt{brandt2000}), and are not included in previous studies of optically selected samples because they can cause an artificial steepening of the $\lx$--$\luv$ correlation.}, which is at present unavoidable, unless UV spectra are obtained for the whole sample. 

The most serious issue potentially affecting a large fraction of the quasar sample is dust extinction. This effect is expected to be stronger at 2500~\AA\ than at 2~keV. Assuming a Galactic dust-to-gas ratio, a relatively small amount of dust, associated with a column density of a few $10^{20}$~cm$^{-2}$, is enough to decrease the observed flux at 2500~\AA~by a factor of $\sim$two, while the flux at 2~keV is nearly unaffected. 

Since dust extinction is expected to redden the observed optical-UV spectrum, in order to analyze this issue we computed for each object the slope $\Gamma_{1}$ of a $\log(\nu)-\log(\nu F\nu)$ power law in the 0.3--1 $\mu$m (rest frame) range, and the analogous slope $\Gamma_{2}$ in the 0.3--0.5 $\mu$m range (rest frame). 
The $\Gamma_{1}-\Gamma_{2}$ distribution is shown in Figure~\ref{fig1}. 
As expected, the two slopes are in general well correlated. The large tail of the $\Gamma_{1}-\Gamma_{2}$ distribution towards low values (i.e. redder spectra) is indicative of possible dust reddening. We assumed a standard SMC extinction law (\citealt{prevot84}, appropriate for unobscured AGN, \citealt{2004AJ....128.1112H,salvato09}) to estimate the $\Gamma_1-\Gamma_2$ correlation as a function of extinction. We obtained the red solid line shown in Figure~\ref{fig1}, where the starting point (with zero extinction, plotted with a star) is derived from the SED of \citet[i.e. $\Gamma_1=0.95$, $\Gamma_2=1.35$]{2006AJ....131.2766R}. This line is \rev{consistent} with the best fit line to the observed points, and confirms that the main physical driver of the large color distribution is dust reddening. This implies that the observed fluxes at 2500~\AA\ are in many cases significantly underestimated.  The dashed lines, orthogonal to the solid one, show the colors corresponding to different values of the ratio $r_{UV}$ between the intrinsic and observed flux at 2500~\AA. The corresponding attenuation at 2~keV, assuming a Galactic dust-to-gas ratio, is instead negligible (about 5\% if the ratio between intrinsic and observed flux at 2500~\AA\ is $r_{UV}=10$).

\begin{figure}
 \resizebox{\hsize}{!}{\includegraphics{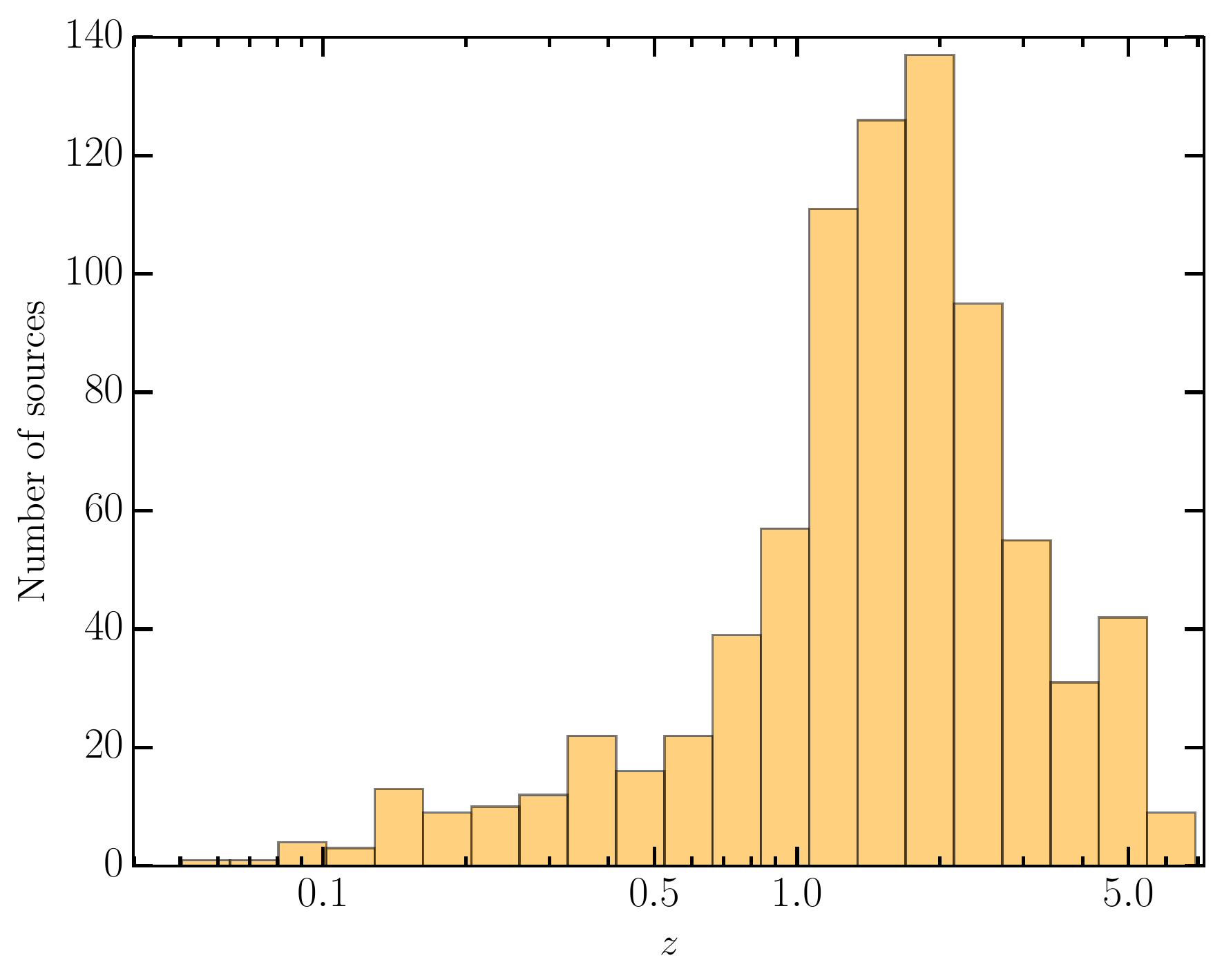}}
\caption{Redshift distribution of the quasar sample used for the cosmological analysis. The total number of sources is 808.  \label{fig2}}
\end{figure}

\begin{figure}
 \resizebox{\hsize}{!}{\includegraphics{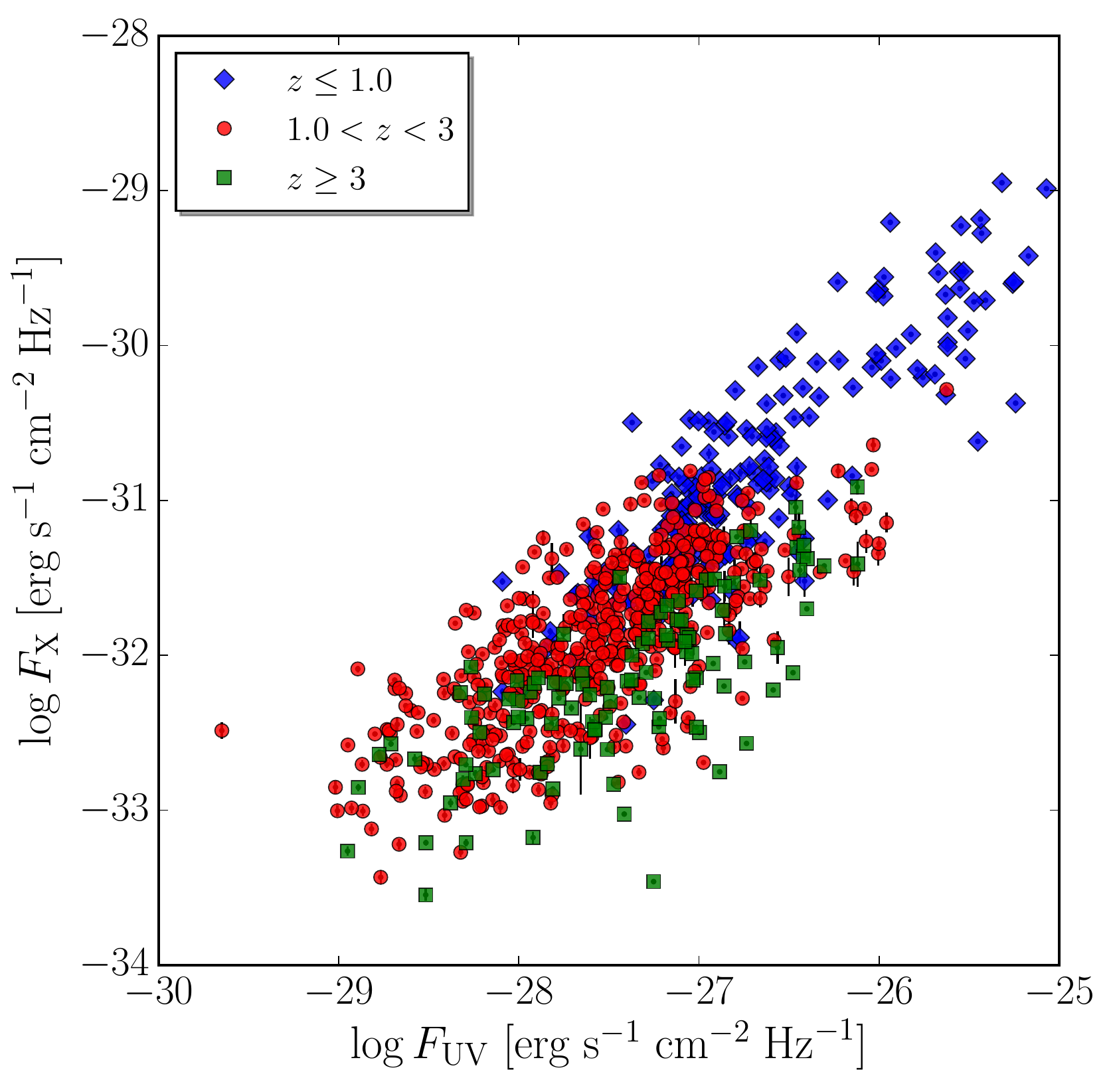}}
\caption{$\fuv-\fx$ distribution for our sample, with colors and point shapes representing different redshift intervals, as indicated in the plot. The correlation between $\fuv$ and $\fx$ is due to the physical correlation between the corresponding luminosities. The observed, large dispersion is due both to the intrinsic dispersion of the $\luv-\lx$ relation and to the redshift distribution at each flux interval. The information on the cosmological parameters is therefore encoded in this dispersion, as visually shown by the three redshift intervals, with the lowest (blue diamonds) and highest (green squares) redshift objects lying preferentially on the upper and lower envelopes of the distribution, respectively.   \label{fig3}}
\end{figure}

In order to obtain an optimal sample for the analysis of $\fx$--$\fuv$ relation, we need to correct the observed UV fluxes for dust extinction. This correction likely introduces further dispersion to the correlation, due to the spread in the intrinsic SED of quasars, and the possible differences in the dust properties (and hence in the extinction law) among quasars. Such uncertainties are likely to be larger when the required extinction correction is also larger. We therefore need to choose an optimal balance between including more objects (so increasing the statistics) and applying smaller extinction corrections (so reducing the dispersion). This choice does not introduce biases in the subsequent analysis, and can be made following the criterion of the smallest errors in the final best fit parameters of our analysis. We therefore repeated the cosmological analysis (discussed in detail the next Section) for several choices of the maximum allowed $r_{UV}$, and we selected the sample with the smallest errors in the determination of $\Omega_M$ and $\Omega_\Lambda$ (specifically, the one minimizing the area of the 1--$\sigma$ $\Omega_M-\Omega_\Lambda$ contour). 
As noted above, absorption effects on the X-ray fluxes are expected to be less important. However, some extreme case are possible, for example if a very low dust to gas ratio implies low UV extinction and relatively high X-ray gas absorption. In these cases the spectrum is much flatter than in standard quasars, with photon indexes $\Gamma<1$. At the other extreme, in a few cases it is possible to observe very steep spectra $\Gamma>3$ due to contamination by galactic sources in galaxies with a very high star formation. In order to exclude these extreme outliers, we required the photon index of our sources to be in the standard range for unobscured, luminous quasars, $\Gamma=1.8\pm1$. 
 

The final ``best sample'' is summarised in Table~\ref{summarysamplecosmo} and it consists of 808 objects with $r_{\rm UV}<10$ (BAL and radio-loud quasars have been neglected). In Figures~\ref{fig2} and \ref{fig3} we plot the redshift and $\fuv-\fx$ distributions. 
Optical and X--ray luminosities/fluxes with their uncertainties are listed in Table~\ref{samplecosmo} (different subsamples are flagged from 1 to 5, see Note in Table~\ref{samplecosmo}).  

\begin{table*}
\centering
\caption{Summary of the best quasar sample. \label{summarysamplecosmo}}
\begin{tabular}{ccccccl}
\hline\hline\noalign{\smallskip}
 Number & $z$ & $\log\luv$ & $\log\lx$ & Reference  \\
              &        & erg sec$^{-1}$ Hz$^{-1}$ & erg sec$^{-1}$ Hz$^{-1}$ &  \\
\noalign{\smallskip}\hline\noalign{\smallskip}
  222         &  $0.061-6.280$ & $28.94-32.90$ & $25.14-28.09$ & \citet{steffen06}  \\    
    20         &  $1.760-4.610$ & $32.14-32.85$ & $27.14-28.07$ & \citet{just07} \\   
    14         &  $4.720-6.220$ & $31.38-32.42$ & $26.62-27.62$ & \citet{shemmer06}  \\ 
  327         &  $0.345-4.255$ & $28.96-31.79$ & $25.08-27.14$ &\citet{lusso2010}  \\
  225         &  $0.173-4.441$ & $29.33-32.40$ & $25.08-27.78$ & \citet{young10}  \\        
\hline\noalign{\smallskip}  
 808         & $0.061-6.280$ & $28.94-32.90$ & $25.08-28.09$ & Total \\     
\hline\hline\noalign{\smallskip}               
\end{tabular}                                                                                 
\end{table*}                                                                    

\begin{table*}
\centering
\caption{Optical and X--ray properties of the ``best sample''. \label{samplecosmo}}
\begin{tabular}{lcccccccc}
\hline\hline\noalign{\smallskip}
 Name & RA & DEC  & Redshift & $\log \luv$ &  $\log \lx$  & $\log \fuv$ & $\log \fx$  & Group$^{\mathrm{a}}$  \\
           &      &    &     & $\rm{[erg\,s^{-1}Hz^{-1}]}$ & $\rm{[erg\,s^{-1}Hz^{-1}]}$ & $\rm{[erg\,s^{-1} cm^{-2}Hz^{-1}]}$ & $\rm{[erg\,s^{-1} cm^{-2}Hz^{-1}]}$ &  \\
\noalign{\smallskip}\hline\noalign{\smallskip}
 022356.30-085707.8  &  35.985  &  -8.952 & 1.575 & 30.92 & 26.67 & -27.29 $\pm$ 0.01 & -31.54 $\pm$ 0.10 & 1 \\
 022435.92-090001.4  &  36.150  &  -9.000  & 1.611 & 30.97 & 26.76 & -27.27 $\pm$ 0.01 & -31.47 $\pm$ 0.11 & 1 \\
 023306.25+004614.5 &  38.276   &   0.771 & 2.292 & 30.66 & 26.96  & -27.94 $\pm$ 0.03 & -31.65  $\pm$ 0.07 & 1 \\
\hline\noalign{\smallskip}                      
\end{tabular}                                   
                                                
\begin{list}{}{Notes---This table is presented in its entirety in the electronic edition; a portion is shown here for guidance.}
\item[$^{\mathrm{a}}$]Group flag: \citet{steffen06} (1), \citet{just07} (2), \citet{shemmer06} (3), XMM--COSMOS (4), \citet{young10} (5).
\end{list}                                      
\end{table*}                                                                    

\section{Analysis of the $\lx-\luv$ relation}
\label{Analysis of the lum relation}
\begin{figure}
 \resizebox{\hsize}{!}{\includegraphics{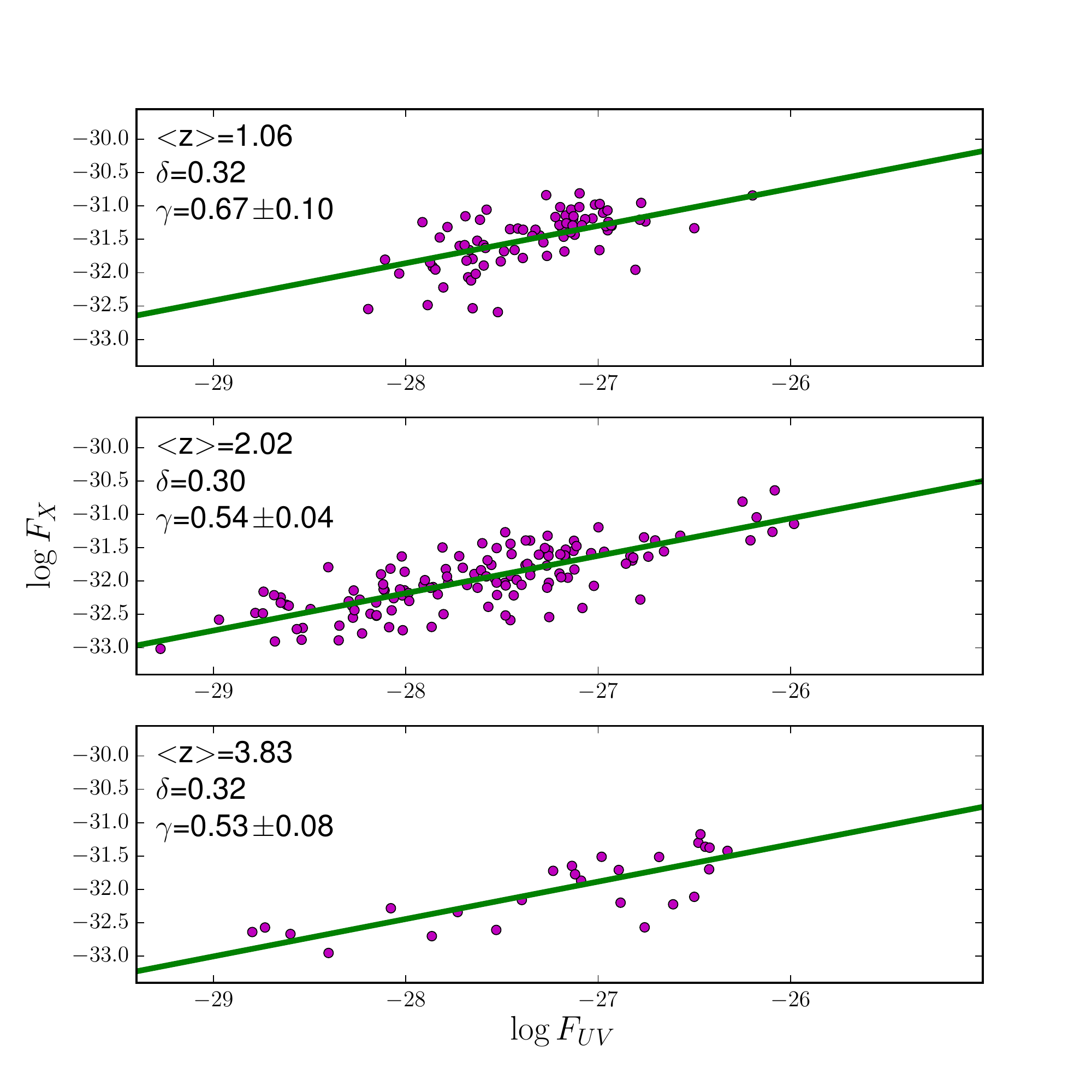}}
 \resizebox{\hsize}{!}{\includegraphics{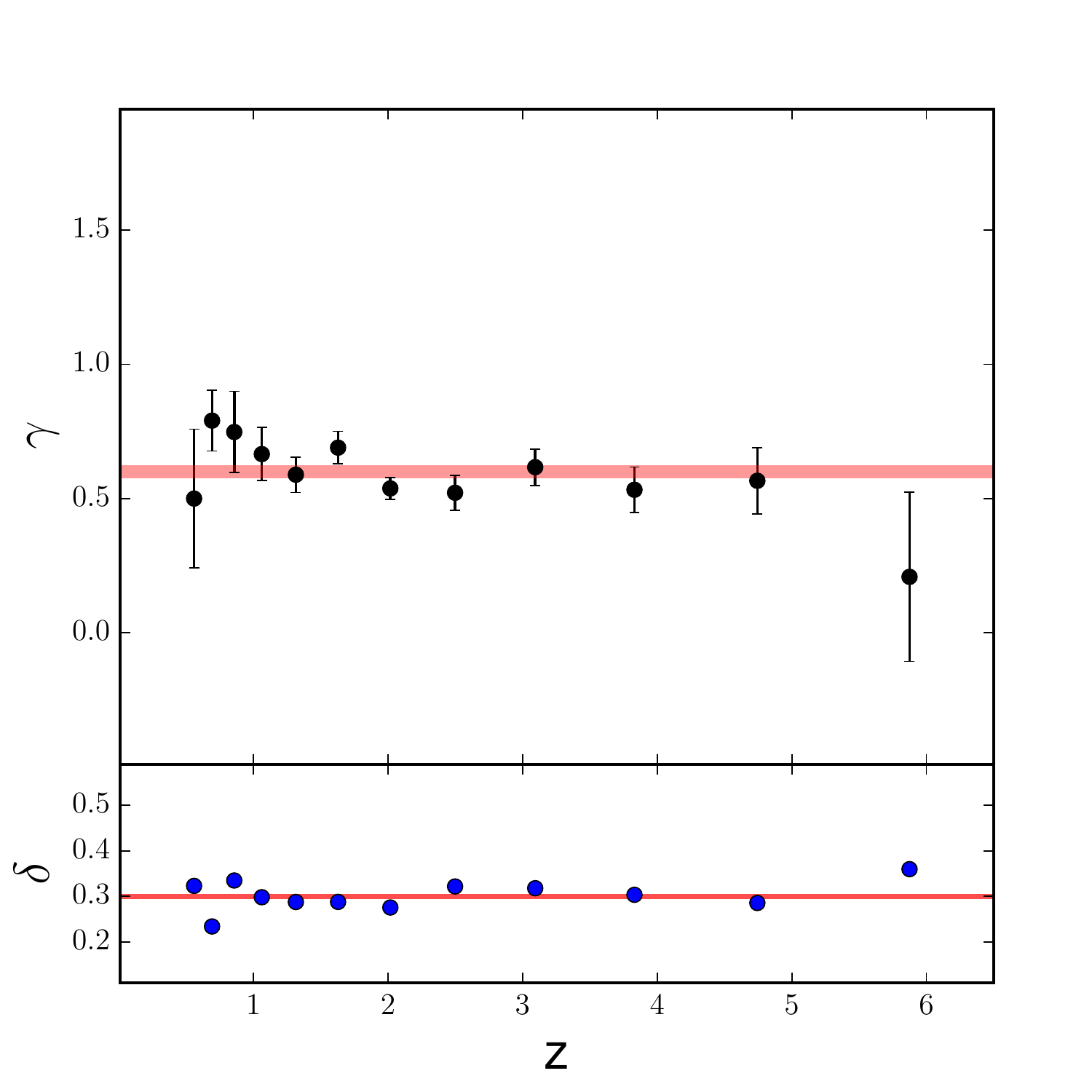}}
\caption{$\fx$--$\fuv$ correlation in narrow redshift intervals. The dispersion due to the different distances within each redshift interval is negligible with respect to the intrinsic dispersion of the $\fx-\fuv$ correlation (see text for details). Therefore, the relation between fluxes is a good proxy of that between luminosities. Upper panel: examples of the correlation in three redshift intervals at $z\sim$1,2 and 4, respectively. Lower panel: best fit values and dispersion $\delta$ of the correlation slope, $\gamma_z$, in each redshift interval. The horizontal lines show the average values, i.e. $\langle\gamma_z\rangle=0.60\pm0.02$ and $\langle\delta\rangle=0.3$. \label{fig4}}
\end{figure}

Based on the method outlined in Section~\ref{Method}, we divided our sample in narrow redshift bins, in order to check the redshift dependence of the $\luv-\lx$ relation. The two requirements for such analysis are (1) that the scatter due to the different luminosity distances within each bin is small compared with the intrinsic dispersion and (2) that each bin is sufficiently populated for a meaningful check. The first condition is met for bins equally spaced in $\log z$, with $\Delta\log z\leq0.1$. The second condition is fulfilled for redshifts $z>0.5$, while for lower redshifts (and narrower intervals, due to the first condition) we would have less than 15 objects in each bin (Figure~2). Therefore, we performed this analysis only in the $z=0.5-6.5$ range. 

We divided the sample in 12 intervals with $\Delta\log z=0.1$, and for each redshift interval we performed a linear fit of $\log \fx-\log \fuv$ relation,  $\log \fx=\gamma_z \log \fuv+\beta_z$, with free $\gamma_z$ and $\beta_z$. 
The results for the parameter $\gamma_z$ are shown in Figure~\ref{fig4}, where we plot the $\gamma_z-z$ relation, and three examples of our fits in three intervals centered at redshift $z\sim$1,2 and 4. 
The value of $\gamma_z$ is consistent with a constant value at all redshifts, without any significant evolution. The average gives $\gamma=\langle\gamma_z\rangle=0.60\pm0.02$. 

\section{Determination of the cosmological parameters}
\label{Determination of the cosmological parameters}

\begin{figure*}
\includegraphics[angle=0,width=1.0\textwidth]{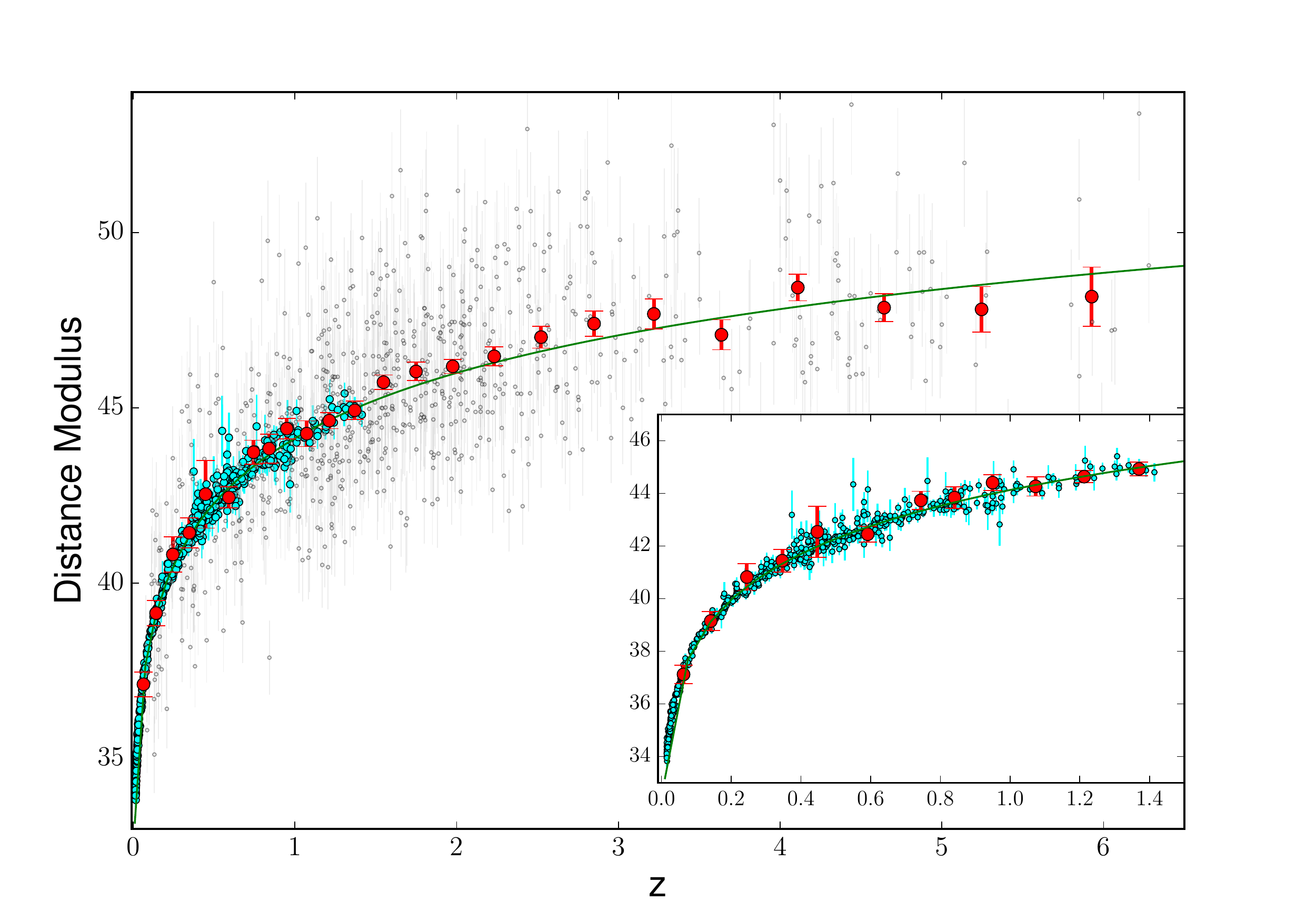}
\caption{Hubble Diagram for the quasar sample (small gray points) and supernovae (cyan points) from the Union~2.1 sample (Suzuki et al.~2012). The large red points are quasar averages in small redshift bins. The inner box shows a zoom of the $z=0-1.5$ range, in order to better visualize the match between the SNe and the quasar samples. The continuous line is obtained from a joint fit of the two samples assuming a standard $\Lambda$CDM cosmological model. The relative normalization between SNe and quasars is a free parameter of the fit, and it is estimated with an uncertainty lower than 1\%.
   \label{hubble}}
\end{figure*}

We adopted the average value of $\gamma$ obtained in the previous Section, $\gamma$=0.60$\pm$0.02, to estimate a distance modulus, $DM=5\log[(D_L/(10~{\rm pc})]$ for each quasar in our sample, defined from Equations~2,~3 as:
\begin{equation}
\label{eq:dm}
DM=\frac{5}{2(\gamma-1)}[\log(\fx)-\gamma\log(\fuv)-\beta'].
\end{equation}
Since the intrinsic value of $\beta$ in not known, $\beta'$ can be considered an arbitrary scaling factor. 

 In Figure~\ref{hubble} we plot our results for each quasar, and for averages in narrow redshift bins (for visual purpose only). We fitted the $DM-z$ relation as discussed in Section~\ref{Method}, assuming a $\Lambda$CDM model, with $\Omega_\Lambda$, $\Omega_M$, the scaling parameter $\beta'$ and an intrinsic dispersion $\delta$ as free parameters.  We also performed a direct fit of $\fx$ as a function of $\fuv$ and $z$, with $\gamma$, $\beta$, $\Omega_\Lambda$, $\Omega_M$, and  $\delta$ as free parameters. The two methods provide fully consistent results (as expected, since they are two different rearrangements of the same data). A summary of the results of cosmological fits is provided in Table~\ref{tbl-cosmo}. 
\rev{ The best fit value of the dispersion is $\delta$=0.30, lower than the values previously reported in the literature. This is a positive effect of our source selection, which removed from the sample many outliers with poor quality data and/or affected by absorption. }

One of the most interesting aspects of the analysis through the distance $DM-z$ relation is that it provides a direct way to compare and merge our data with those from supernovae (SNe). We can of course compare our constraints with those from any cosmological measurement, such as CMB, BAO, clusters, etc. However, this analysis can be done only on the final values of the cosmological parameters obtained from each method independently, since each method is based on the measurement of different physical quantities (CMB fluctuations, baryonic oscillations, etc.). In our case instead, both SNe and quasars are used as ``standardized candles'' and the cosmological parameters are obtained from the same physical quantity, i.e. the distance modulus. Technically, the only difference is the absolute calibration, or the ``zero-point'' of the $DM-z$ relation, which can be measured for SNe, but is unknown for quasars. As a consequence, a joint fit of the $DM-z$ relation for SNe and quasars requires that the $\beta'$ parameter in Equation~\ref{eq:fuvfx} is used as a free parameter of the fit in order to cross-calibrate the $DM$ of the two group of sources.

\begin{figure}
\includegraphics[angle=0,width=0.5\textwidth]{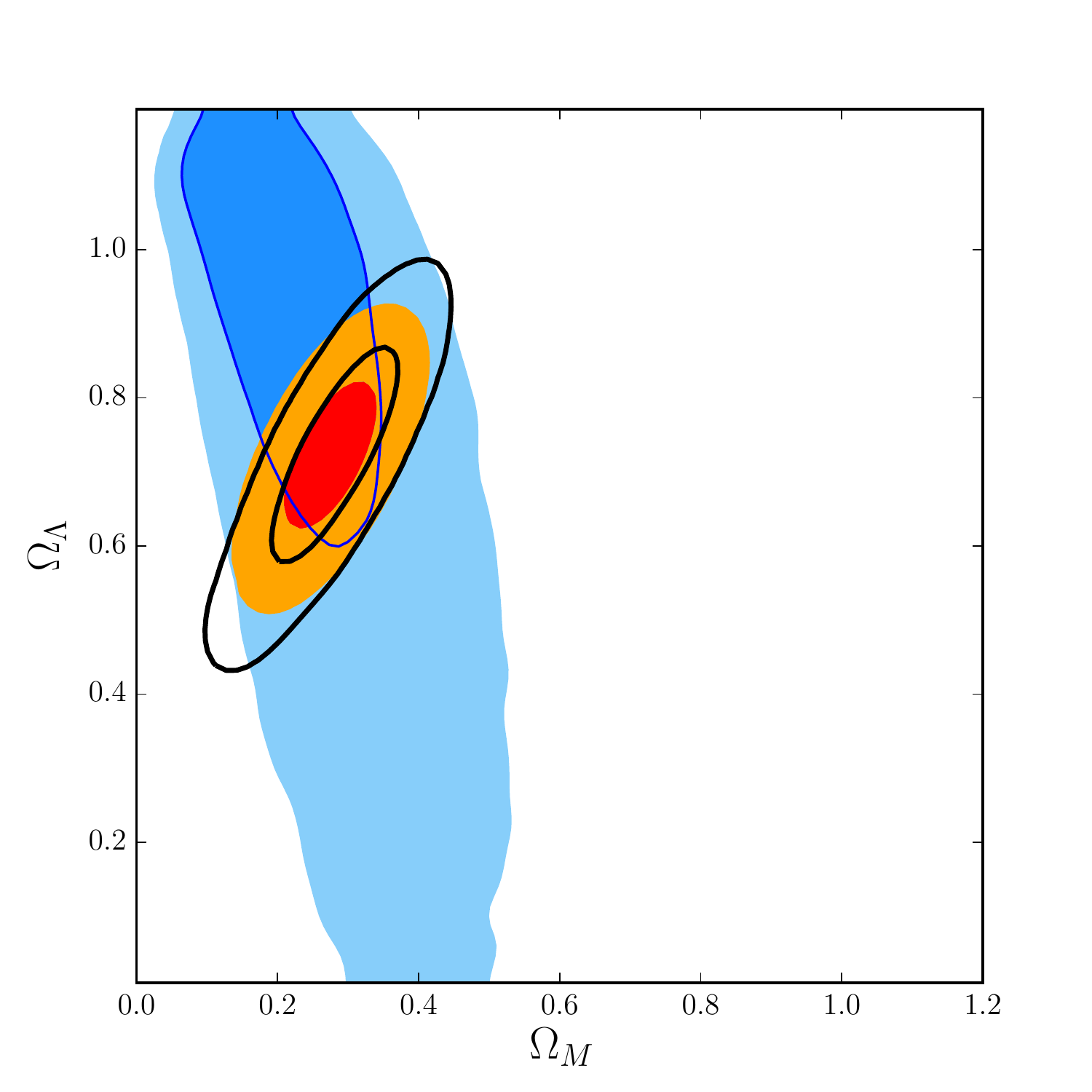}
\caption{68\% and 95\% contours for $\Omega_M$ and $\Omega_\Lambda$, assuming a standard $\Lambda$CDM model, as derived from our analysis of the Hubble diagram of quasars (blue), from the SNe Union 2.1 sample \citet{suzuki12} (empty black), and from a joint fit (orange-red) \label{fagiolo}}
\end{figure}

For our joint quasar-SNe joint analysis we used the Union~2.1 SNe sample of the Supernovae Cosmology project\footnote{http://supernova.lbl.gov/Union/} (Suzuki et al.~2012). We derived the cosmological parameters from the Hubble diagram of SNe only (obtaining the same results as in Suzuki et al.~2012, and so demonstrating the full consistency of our analysis with the published one) and for the joint Hubble Diagram. 
The results for the determination of the cosmological parameters $\Omega_M$ and $\Omega_\Lambda$ are shown in Figure~\ref{fagiolo}. The 68\% and 95\% contours for $\Omega_M$ and $\Omega_\Lambda$, assuming a standard $\Lambda$CDM model, are derived from our analysis of the Hubble diagram of quasars only (blue contours), from the SNe Union 2.1 sample \citet[empty black contours]{suzuki12}, and from a joint fit quasars + SNe (orange-red contours). Thanks to the large redshift overlap between the SNe and quasars samples (inner box of Figure~5) the cross-calibration parameter $\beta$' is estimated with an uncertainty lower than 1\%.
We note that the contour plot relative to the joint sample is not a statistical intersection of the two individual contours, but the result of a simultaneous fit of the $DM-z$ relation for the combined sample. 

\begin{table*}
\begin{center}
\caption{Results of cosmological fits.\label{tbl-cosmo}}
\begin{tabular}{lccccccc}
\hline\hline\noalign{\smallskip}
           &  \multicolumn{2}{c}{$\Lambda$CDM\tablenotemark{b}} & \multicolumn{2}{c}{$w$CDM\tablenotemark{c}} & \multicolumn{3}{c}{$w_0$$w_a$CDM\tablenotemark{d}}\\
           &  $\Omega_M$ & $\Omega_\Lambda$ &  $\Omega_M$ & $w$ & $\Omega_M$ & $w_0$ & $w_a$ \\
\tableline\noalign{\smallskip}
DATA\tablenotemark{a}           &  0.22$^{+0.10}_{-0.08}$ & 0.92$^{+0.18}_{-0.30}$ &             &   &       &   &    \\
DATA\tablenotemark{a}+SNe       &  0.28$^{+0.04}_{-0.04}$ & 0.73$^{+0.08}_{-0.08}$ &             &   &       &   &    \\
10$^5$ QSO\tablenotemark{a}     &  0.27$^{+0.02}_{-0.02}$ & 0.69$^{+0.08}_{-0.07}$ & 0.26$^{+0.02}_{-0.015}$ & -0.90$^{+0.15}_{-0.14}$ &      &    &   \\
10$^5$ QSO\tablenotemark{a}+SNe &  0.27$^{+0.02}_{-0.02}$ & 0.70$^{+0.04}_{-0.04}$ & 0.26$^{+0.02}_{-0.014}$ & -0.96$^{+0.05}_{-0.05}$ &       &   &    \\
10$^6$ QSO\tablenotemark{a}     &  0.27$^{+0.02}_{-0.01}$ & 0.73$^{+0.01}_{-0.01}$ & 0.27$^{+0.02}_{-0.01}$ & -1.02$^{+0.02}_{-0.02}$ & 0.27$^{+0.03}_{-0.02}$ & -1.03$^{+0.05}_{-0.05}$  & 0.3$^{+0.4}_{-0.4}$       \\
\tableline\noalign{\smallskip}
\end{tabular}
\tablecomments{$a$: Data samples: DATA: our actual sample; 10$^5$~QSO: the simulated 10$^5$ QSO sample from the cross correlation between the SDSS DR7 and DR10 quasar catalogs and the eROSITA all-sky survey; 10$^6$~QSO: simulated sample of one million quasars from future all sky surveys (see text for details).
$b$: Standard $\Lambda$CDM model.
$c$: Cosmological model assuming a flat Universe and a non-evolving dark energy parameter $w$.
$d$: Cosmological model assuming a flat Universe and an evolving dark energy equation of state: $w=w_0+w_a\times(1+z)$.}
\end{center}
\end{table*}

\section{Discussion}
\label{Discussion}

The main results of our work are the validation of the linear  $\log \luv-\log \lx$ relation at all redshifts, and its application as a cosmological probe.
The physical origin of the relation is unknown. Here we only notice three relevant aspects.
\begin{enumerate}
 \item The analysis in narrow redshift bins has shown no significant redshift evolution of the shape of the relation (a linear fit is adequate at all redshifts) and of the correlation slope, $\gamma$. We cannot directly test the constancy of the parameter $\beta$, but an important consistency check of this point is provided by the comparison between the Hubble diagrams of quasars and SNe at $z<1.4$ (the inner panel in Figure~\ref{hubble}). The perfect match between the two curves is indicative that the evolution of the $\beta$ parameter is negligible. Either that, or there is some unknown redshift evolution is also present between luminosity and redshift in SNe, with the same dependence as for the $\log \luv-\log \lx$ for quasars. The need to invoke this ``physical conspiracy'' is enough to rule out this possibility. Formally, the possibility of a redshift evolution of the  $\log \luv-\log \lx$ relation beyond $z=1.4$ remains, even if it appears quite unlikely, and not based on any physical motivation. On this regard, we notice that despite a large dispersion in many observational properties of quasars, their optical-UV SED does not show any evolution with redshift and/or luminosity, up to the most extreme cases, as shown, for example by the similarity between the average quasar SED \citep{2006AJ....131.2766R,2014MNRAS.438.1288H} and the SED of a luminous $z\sim7$ quasar \citep{2012MNRAS.419..390M}. The same absence of evolutionary/luminosity effects is observed in the spectral properties of quasars in hard X--rays (e.g. \citealt{young10,lusso2010}).
 
\item The cosmological application presented in this paper is not directly based on the physical interpretation of the $\luv-\lx$ relation, but only on its observational evidence. This situation is qualitatively similar to what is presently known about supernovae as cosmological tools: the ``standardization'' of the supernovae luminosity is obtained through the empirical relation between the peak luminosity and the slope of the luminosity light curve after the peak \citep{1993ApJ...413L.105P}. 

\item The large observed dispersion in the $\luv-\lx$ relation ($\sim0.30$ in logarithmic units, i.e. a factor of two in physical units) is obviously the main limitation for precise cosmological measurements. We note for example that the dispersion in the distance modulus-redshift relation is much smaller for supernovae than for quasars. In other words, quasars as standard candles are much less precise than supernovae. However, the main advantages in using quasars are the wider redshift range, and the possibility of future large improvements of  the sample. This is discussed in detail in the next Section. 
\end{enumerate}
The origin of the intrinsic dispersion must be related to the physical origin of the relation, and is therefore unknown. However, the {\it observed} dispersion may be significantly larger than the intrinsic one. Here we briefly discuss possible contributions to the observed dispersion:

- Inclination effects: assuming a disk-like UV emission and an isotropic X--ray emission, we expect that edge-on objects have a relatively low observed UV-X--ray luminosity ratio \citep{2011MNRAS.411.2223R}. This implies that edge-on quasars are outliers in the $\log \luv-\log \lx$ relation, and a contribution to the dispersion comes from the distribution of disk inclinations. In an optical-UV selected sample the distribution of inclination angles $\theta$ has a peak value  at $\cos\theta=5/9$, with a dispersion $\sigma(\cos\theta)=2/9$ \citep{2011MNRAS.411.2223R}. On average, this $\sim$40\% contribution to the dispersion is negligible with respect to the total one (after correction, the observed dispersion would decrease from 0.30 to $\sim$0.29), but this effect may be responsible for some of the largest outliers. In an X--ray selected sample, there would be no selection bias with respect to the disk inclination (assuming isotropic X--ray emission), and the dispersion due to the inclination effect would be slightly larger, but still negligible with respect to the total one.

- Variability: our optical-UV and X--ray observations are not simultaneous, therefore variability must add a significant contribution to the observed dispersion. Previous studies of \citet{2008ApJ...685..773G} on SDSS quasars suggest that variability cannot explain all of the observed dispersion in the $\alphaox-\luv$ relation. It is however possible that simultaneous observations (for example, taking advantage of the Optical Monitor on-board XMM-{\it Newton}, and/or the XRT and UVOT instruments on-board {\it Swift}) could reduce the intrinsic dispersion, as suggested by \citet{vagnetti2010,vagnetti13}. We note however that even if variability is responsible for a significant fraction of the observed dispersion, simultaneous observations may be not enough to obtain a tighter correlation, if the delays between optical and X--ray variations are longer than the typical observing time. 
This seems to be the case in low-luminosity AGN, such as NGC~5548 \citep{Edelson2015} where the X--ray emission leads the optical one by about a day. It is reasonable that in higher luminosity quasars the delays are significantly longer. 

- Absorption correction. We corrected our sample as explained in Section~3, assuming an intrinsic quasar SED as in Richards et al.~(2006), and a standard extinction law. This simple approach has several limitations:\\
1) We neglected possible contributions from the host galaxy. While this is likely a safe assumption for the X--ray emission, it is possible that the contribution of the galaxy in the optical--UV cannot be ignored, especially for low luminous AGN ($L_{BOL}\leq10^{44}$ erg s$^{-1}$). This would alter both the estimate of the UV flux of the quasar, and the absorption correction. We expect this contribution not to be relevant for most objects, given the high luminosity of our sources (average $\log(L_{BOL})$=46.3\footnote{Estimated from the luminosity at 2500~\AA\ and assuming a bolometric correction of 3, as in Krawczyk et al. (2013).}), and the good match between the $\Gamma_1-\Gamma_2$ distribution with the expectations from dust extinction (Figure~1). \\
(2) We assumed the same SED for all sources, while intrinsic differences are present even in blue color/UV-selected samples (Elvis et al.~1994, Richards et al.~2006).\\
(3) We assumed the same dust extinction law for all sources, neglecting possible differences in the composition of dust grains in each object. \\
All these simplifications likely contribute to the observed dispersion, which may be reduced with a future, more detailed analysis of the optical--UV SED of each source.

- Systematic effects. A final concern regarding our findings is related to possible systematic effects in the sample selection. 
Specifically, the fits of the Hubble diagram may be biased in different ways if the observations in one of the two bands are biased towards high values. In our sample, such a bias may be present due to our choice to include only sources with both UV and X-ray measurements, avoiding upper limits. The Steffen et al.~(2006) and Young et al.~(2010) subsamples, both optically selected, contain a small fraction of X-ray non-detections (12\% and 13\%, respectively). Analogously, the Lusso et al.~(2010) X--ray selected subsample could  be undersampled at low UV fluxes. In the latter case the effect on the fit of the F$_X$--F$_{UV}$ relation is expected to be smaller than for UV-selected samples, due to the broader flux range in the optical-UV than in X-rays (Figure~\ref{fig2}). In order to further investigate this issue, we simulated the effect of cutting the available sample at a relatively high X--ray flux values ($\log(F_{\rm X,min})=-32$ in the same units as in Figure~\ref{fig2}). This is equivalent to assuming that our sample is derived from the cross correlation between an optically selected catalog and an X--ray survey with a uniform X--ray lower limit $F_{\rm X,min}$. The results are a significantly lower slope of the $\log\luv-\log\lx$ relation ($\gamma\sim0.5$), and a distorted Hubble diagram, providing cosmological parameters biased towards lower values of both $\Omega_M$ and $\Omega_\Lambda$. Making a general statement from the result of this test is not straightforward, since the effects of such biases on the estimates of the cosmological parameters depend on both the redshift and flux distributions of the sample. Different redshift distributions may have a different impact of the flux limit on the Hubble diagram. 
It is also difficult to quantitatively estimate the possible effects of these kinds of biases on our findings, given the non-homogeneous composition of our sample. However, we are confident about our results, based on the following two considerations:
\begin{enumerate}
\item  The results of the analysis of the $\gamma_{OX}$--L$_{UV}$ relation in Young et al.~(2010) are insensitive to the inclusion/exclusion of upper limits, suggesting that the bias is small. \\
\item We checked for the possible effects of these biases by cutting our sample at different UV and X--ray minimum fluxes, starting from $\log(F_{\rm X,min})=-33.5$ and $\log(F_{\rm UV,min})=-29$. \rev{ We already discussed the case with a cut at  $\log(F_{\rm X,min})=-32$, where the effect of the bias was not negligible. However, we verified that the results are not significantly altered up to minimum fluxes of $\log(F_{\rm X,min})\sim-32.5$.} This shows that our current sample is not seriously affected by systematics related to flux limits. We note that this may become a much more serious issue with future, larger samples which will allow more precise measurements of the cosmological parameters, and will therefore be sensitive to smaller systematic effects.  
\end{enumerate}

\section{Future developments}
\label{Future developments}

\begin{figure*}
\plottwo{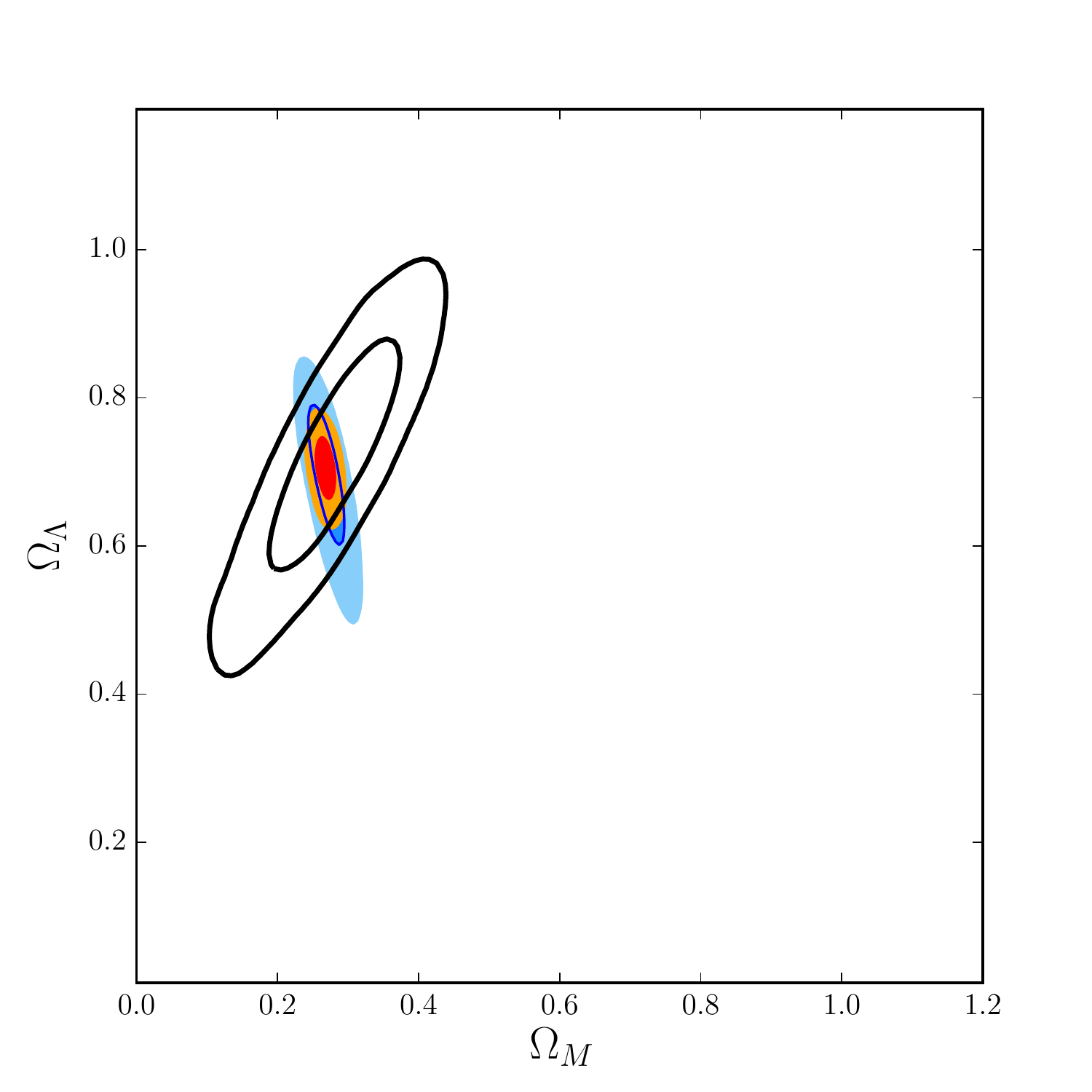}{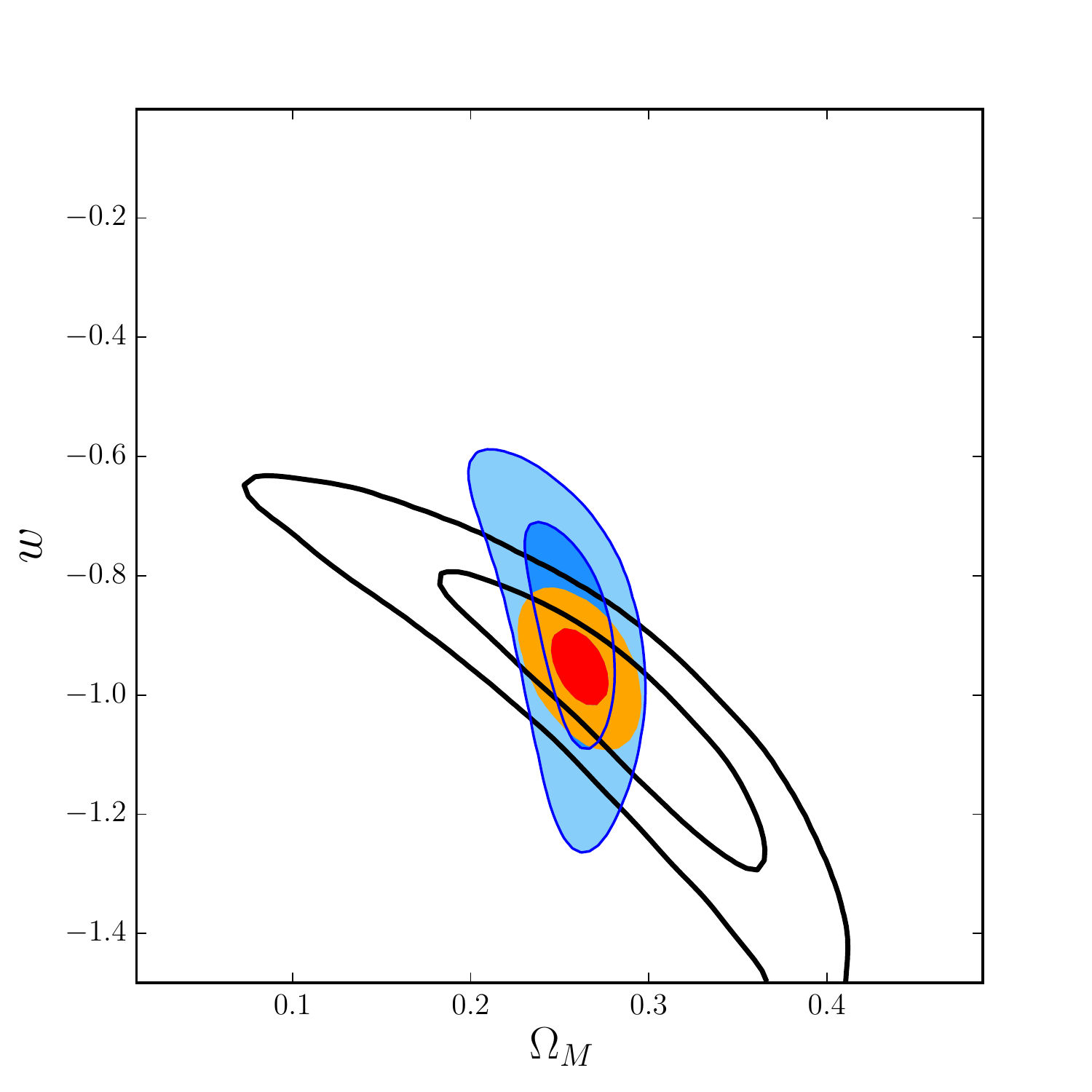}
\caption{Simulations for a sample of 100,000 quasars obtained cross-correlating the SDSS DR7+DR10 quasar catalogs with the future eROSITA X--ray all-sky survey (see text for details). Left panel: 68\% and 95\% contours for $\Omega_M$ and $\Omega_\Lambda$ in a $\Lambda$CDM model (blue: quasars; empty black: SNe Union 2.1 sample \citet{suzuki12}; orange-red: joint fit). Right panel: same for a model with the dark energy equation of state parameter $w$, assumed to be constant at all redshifts ($w=-1$ is equivalent to having a cosmological constant). \label{sim50000}}
\end{figure*}

\begin{figure*}
\includegraphics[width=0.33\textwidth]{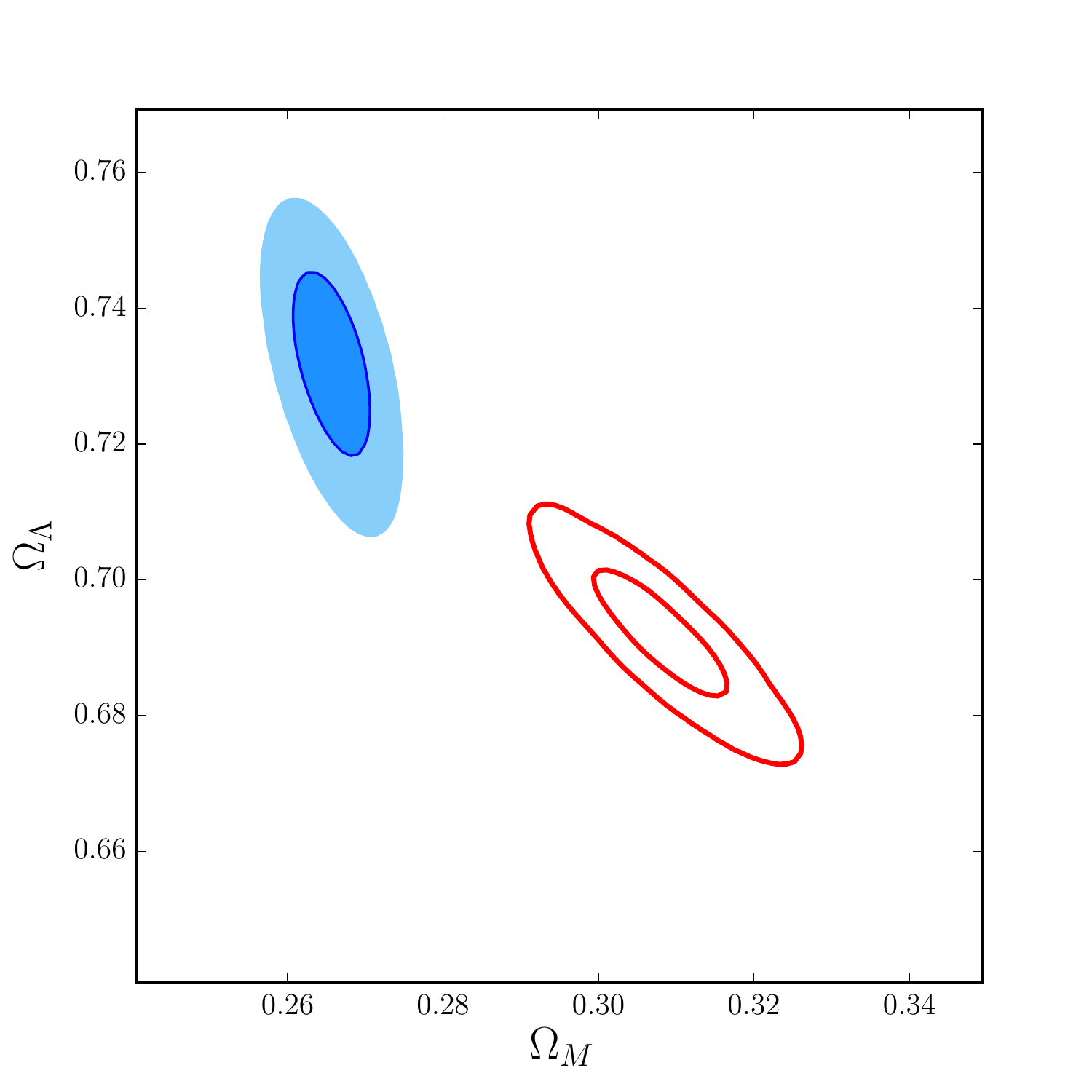}
\includegraphics[width=0.33\textwidth]{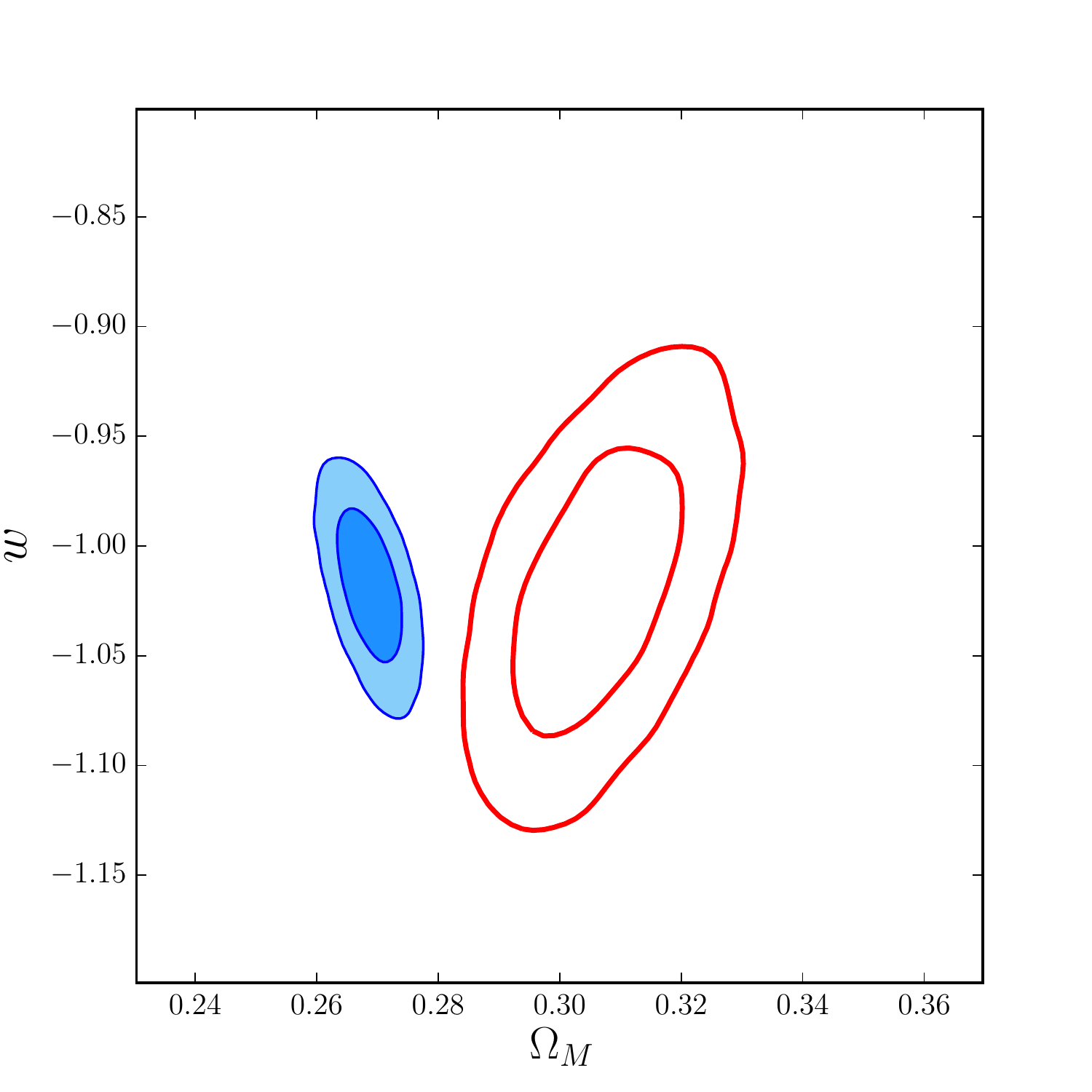}
\includegraphics[width=0.33\textwidth]{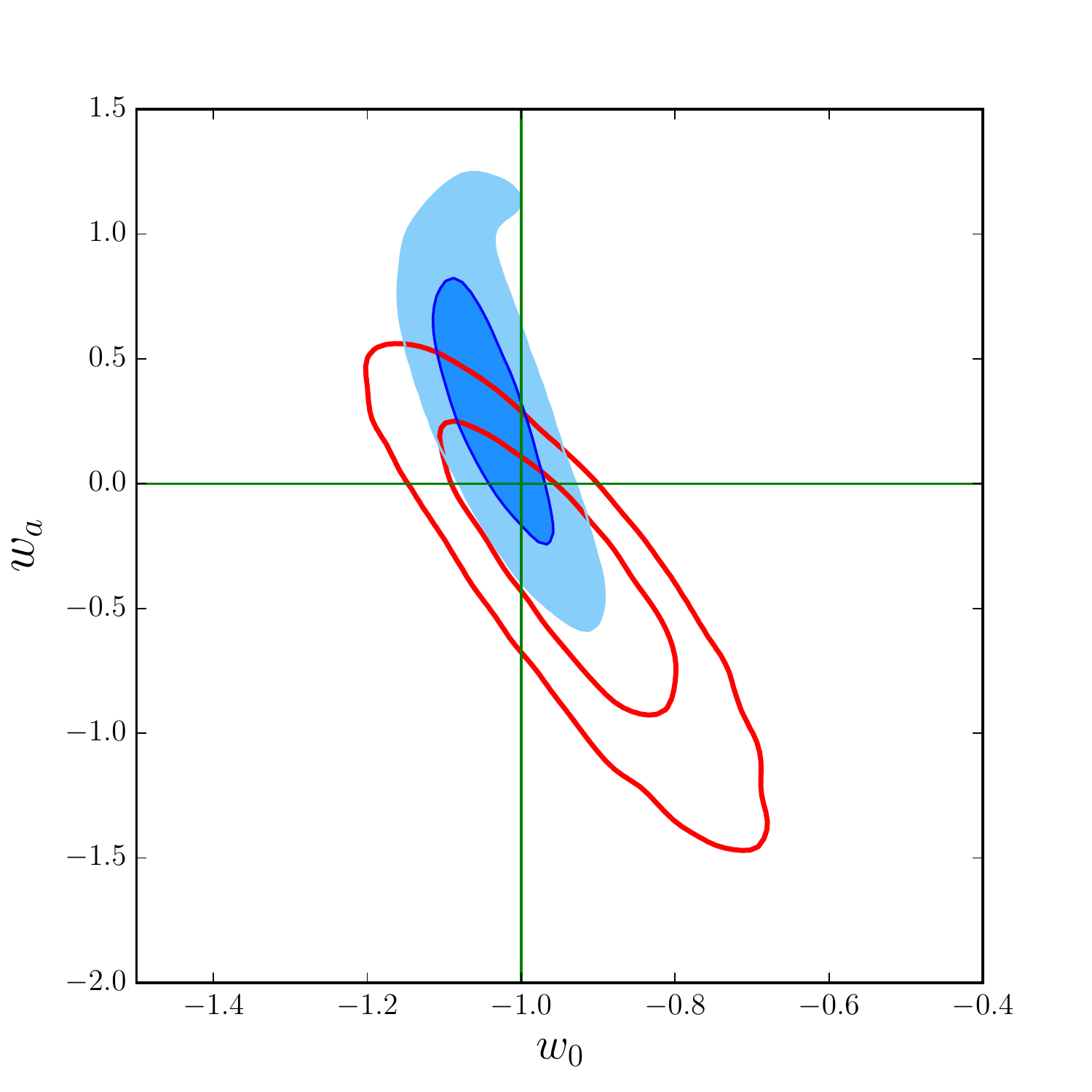}
\caption{Simulations for a sample of 1,000,000 quasars with the same $\fuv-z$ distribution as our present sample (blue contours), compared with the best measurements available today, obtained from Planck, lensing, BAO and supernovae measurements \citep[empty red contours]{2015arXiv150201589P}. Left panel: 68\% and 95\% contours for $\Omega_M$ and $\Omega_\Lambda$ in a $\Lambda$CDM model; Middle panel: same for a model of a flat Universe with the dark energy equation of state parameter $w$, assumed to be constant at all redshifts ($w=-1$ is equivalent to having a cosmological constant). Right panel: same for a model of a flat Universe with evolving dark energy equation of state, $w=w_0+(1+z) w_a$. The mismatch between the contours are due to the difference between the simulated values ($\Lambda$CDM model with $\Omega_M=0.27$ and $\Omega_\Lambda=0.73$) and the best estimates from the Planck+BAO+lensing+SNe measurements \citep[$\Omega_M\sim0.31, \Omega_\Lambda\sim$0.69]{2015arXiv150201589P}. \label{sim1000000}}
\end{figure*}

The analysis presented in this paper already provided important results: the validation of the $\luv-\lx$ relation at all redshifts, and the extension of the Hubble Diagram up to $z\sim6$. However, the constraints on the cosmological parameters are still loose compared with other methods, and the improvement on the measurement errors obtained by combining our constraints to those from SNe, CMB, BAOs and clusters is not significant. However, the present work is mainly intended as a demonstration of the power of the method, while most of its potential is still to be exploited in future work, both with available and forthcoming data. Here we list the main foreseeable future developments.\\
\subsection{ Sample improvements - present.} The work presented here is based on the data available in the literature, from samples already used to study the $\alphaox-\luv$ correlation. We plan to extend the analysis in at least three directions.
\begin{enumerate}
\item We will increase the sample size at low redshift. The present shape of the $\Omega_M-\Omega_\Lambda$ contour, elongated along the $\Omega_\Lambda$ direction, is due to the poor sampling at low redshift (10\% of the sample is at $z\leq0.5$, only 6 objects are at $z\leq0.1$), where the effect of $\Omega_\Lambda$ is higher. \rev{ Our simulations show that} an optimal redshift distribution of the quasar sample should be about constant in $\log z$. An increase of the number of  sources at low redshift can be obtained by adding local bright Seyfert galaxies and quasars with X--ray and optical-UV observations. The crucial point for nearby and/or low luminosity sources, which requires a source by source check, is the accurate determination of the flux at 2500~\AA\ taking into account for possible galactic contamination, dust extinction, aperture effects. We expect to add about 100 sources from different catalogs in the literature at $z<0.1$. We note that this addition will be important in order to improve the precision in the estimate of the cosmological parameters with quasars {\em alone}, while the constraints from the total Hubble diagram of SNe and quasars will remain unchanged. Indeed, the constraints from SNe are at present much more stringent in the overlapping part of the diagram (i.e. at $z<1.4$). Therefore, the quasars at $z<1.4$ are useful (a) to test the method and (b) to cross-calibrate the SNe and quasar diagrams (we recall that this is needed because the scaling constant $\beta$ of the $\lx$--$\luv$ correlation in quasars in unknown). Since our sample already provides an excellent cross-calibration at $z<1.4$, only new sources above $z=1.4$ will contribute to increase the measurement precision if SNe and quasars are used together. 
\item We will analyze through simulations the effect of adding to the sample bright, intrinsically blue quasars at high redshift ($z>4$) from the SDSS surveys, and propose them for new X--ray observations. This will greatly improve the quality of the Hubble diagram in this redshift interval, and will provide a precise test of the cosmological model in a redshift range where no other method is available. 
\item We will cross-correlate the SDSS quasar catalogs with the {\em XMM-Newton} and {\em Chandra} archives. We expect to obtain several hundred sources with optical-UV and X--ray coverage, which will greatly improve our statistics. Within this project, we plan to compare the results obtained using the SDSS photometry with those from the Optical Monitor \citep{2012MNRAS.426..903P} on-board XMM-{\em Newton}, which are simultaneous with the X--ray observations. This test will be important to investigate the effect of variability on the observed dispersion of the $\luv-\lx$ correlation.
\end{enumerate}
\subsection{Sample improvements - near future.} The final SDSS-III quasar catalog \citep{alam2015} contains at present almost 300,000 objects. For all these sources we can obtain a 2500~\AA\ flux, and a reddening/extinction estimate, as done for the present sample. A major increase in the number of X--ray observations of quasars will be crucial for the future use of quasars as cosmological probes. 
The eROSITA (extended ROentgen Survey with an Imaging Telescope Array) satellite \citep{2012arXiv1209.3114M} is expected to survey the whole X--ray sky down to $\sim10^{-14}~\fluxunits$ ($\sim$2500 quasars over 18 deg$^2$, i.e. $\sim140/$deg$^2$). 
 The fraction of Sloan quasars detected at the eROSITA limiting flux is $\sim25\%$: essentially $1/4$ of all SDSS quasars are expected to be detected by eROSITA (Menzel et al. 2015, MNRAS submitted). 
We applied the same color selection as in Figure~\ref{fig1} to the whole SDSS DR7+DR10 samples, and simulated an X--ray flux measurement for each object.  
We assumed the observed linear $\log \lx-\log \luv$ relation, with slope $\gamma=0.60$ and dispersion $\delta=0.30$, and a $\Lambda$CDM cosmology with $\Omega_M=0.27$ and $\Omega_\Lambda=0.73$. Finally, we conservatively assumed that about half of the SDSS quasar will have an eROSITA X--ray flux, obtaining a final sample of $\sim$100,000 quasars (over the whole area covered by SDSS and eROSITA, i.e. 8,000 deg$^2$). We then fitted to this sample a $\Lambda$CDM model, and a model with a dark energy equation of state parameter $w$ (assuming a flat Universe). The results are shown in Figure~\ref{sim50000}, and demonstrate the potential of quasars as cosmological probes: the precision that can be achieved with SNe and quasars (i.e., only using a Hubble diagram of ``standard candles") is similar to that obtained today by combining all the available methods (see, e.g. Figure~6 of \citealt{suzuki12}, Figure~16 of \citealt{betoule14}). \\
The simulation discussed here assumes only statistical errors, thus neglecting the possible systematic effects due to contribution of emission lines to $\fuv$, to extinction correction, and to the flux limits in the sample selection. We already discussed these points for the present sample, concluding that their effects are negligible. However, the situation may be different with a much larger sample and more precise estimates of the cosmological parameters. In order to further investigate these issues, we performed the following checks:\\
1) We simulated the X--ray fluxes for the sample of SDSS-DR7 quasars of Shen et al. (2010), starting from their estimates of $\fuv$. We then replaced the $\fuv$ values with the ones obtained applying our method based on photometric points, and performed the cosmological analysis. We obtained cosmological parameters significantly shifted with respect to the ones assumed in the simulation ($\Omega_M\sim0.15\pm0.03$, to be compared to the simulated value of 0.27, while $\Omega_\Lambda$ is marginally compatible with the simulated value of 0.73). If a correction like the one discussed in Appendix~\ref{A.4} is applied, the discrepancy is reduced, but is still present. We conclude that, in order to achieve the precision shown in Figure~\ref{sim50000}, a careful spectral analysis is needed. This problem may be reduced by a more detailed SED fitting based on the available SDSS photometric points complemented by NIR (when available) and WISE data. We will further investigate this point in forthcoming papers. However, in the specific case of the SDSS+eROSITA quasar sample, we will be able to precisely estimate $\fuv$ and $\fx$ from an analysis of the UV and X-ray spectra, which will be available for all the sources. An analogous check on the possible bias introduced by the extinction correction is less straightforward, because it is not clear whether the uncertainty in such correction is redshift-dependent. Here we only notice that the statistics will be enough to restrict the sample to the bluer objects, and that a more precise correction can be obtained from a complete spectral analysis rather than from the photometric colors.\\
2) A similar issue as discussed above may be present in the determination of the X-ray fluxes. A simulation analogous to the previous one has been done based on the data in Fig.~9 (right panel, see discussion in Appendix~\ref{A.5}). In this case however we do not find any significant shift in the best fit values of the cosmological parameters. Furthermore, we expect to be able to obtain the value of $\fx$ from a complete spectral analysis for all the sources in the sample.\\
3) The flux limits of our sample may introduce a bias in the observed $L_X$-$L_{UV}$ relation: objects with an expected (based on the relation) flux near to the sample flux limit will be observed only in case of positive fluctuations. This issue is expected to be relevant mostly in the X--rays, due to the larger flux range in the UV than in the X--rays (which in turn is due to the slope of the correlation being $<1$).  There are two ways to deal with this issue: the first one is to include non-detections in the analysis. This will require a different statistical treatment, in order to properly account for censored data. A second way to obtain an (almost) unbiased sample is to include only objects that would be observed even in case of negative flux fluctuations. To do so, for each $\fuv$ we considered the expected 2$\sigma$ lower limit of $\fx$, based on the distribution in Fig.~3. We then included the object in the sample only if this minimum X--ray flux is above the detection limit (regardless of the {\em observed} value of $\fx$). We checked the effects of this cut varying the rejection threshold, and we found that, considering the expected X--ray flux limit of the eROSITA all-sky survey, we may have a bias in the estimate of the correlation slope $\gamma$, which in turn alters the shape of the Hubble Diagram. We found that a filter like the one presented above, with a 2$\sigma$ threshold, is enough to remove the bias. The fraction of rejected objects is of the order of 20\%. Considering that the total number of SDSS quasars is close to 300,000, even applying all the cuts discussed above,  the number of quasars in the final, clean sample will easily remain higher than the value of 10$^5$ we assumed in the cosmological simulations.
\subsection{Sample improvements - in 10--15 years.} We can extend our simulations trying to predict the available data provided by the next generation of major observatories. For example, the ESA mission Euclid will identify a few million quasars through slitless near-IR spectroscopy \citep{2012SPIE.8442E..0TL}, with thousands at very high redshifts ($z>6$); the Large Synoptic Survey Telescope \citep[LSST]{2008arXiv0805.2366I} will discover through photometry and variability millions of quasars, with an efficiency and completeness analogous to that of spectroscopic surveys \citep{2012ApJ...751...52E}. On the X--ray side, the {\em Athena} (Advanced Telescope for High ENergy Astrophysics) observatory will have a large field of view imager, with resolution of a few arcsec, capable of detecting unabsorbed quasars up to $z\sim8$ \citep{2013arXiv1306.2325A}. It is expected that a significant fraction of the observing time of {\em Athena} will be used to perform wide area surveys. The match between these two observatories will likely provide samples of at least several hundred thousands quasars with both optical/UV and X--ray measurements. Even better, X--ray survey telescopes, such as the Wide-Field X--ray Telescope \citep{2009astro2010S.217M} would further increase these numbers by an order of magnitude: about 10$^7$ quasars are expected in the WFXT  surveys, with $\sim$1,600 objectw at z$>$6 \citep{gilli2010}. In order to have a first hint of the possible use of such samples as a cosmological estimator, we simulated a sample of a million quasars, with the same properties as the presently available one. The results are illustrated in Figure~\ref{sim1000000}, and show how the Hubble diagram of quasars may become a fundamental tool for precision cosmology, and in particular for the determination of possible deviations of the dark energy component from the standard cosmological constant. 
We note that this simulation is optimistic in the sense of assuming the availability of these new observational facilities, but is quite conservative in other respects. A million quasars in probably a large underestimate if {\it Euclid}, LSST, and a large area X--ray surveyor all become available. The simulated redshift distribution (obtained from the presently available sample) does not include the thousands of quasars at redshift $z>6$ expected from {\it Euclid}, which would in particular improve the precision of the determination of $\Omega_M$ (which, in turn, is partly degenerate with the $w_a$ parameter in models with an evolving equation of state of the dark energy, $w=w_0+w_a\times z/(1+z)$).

\section{Conclusions}
\label{Conclusions}
We have demonstrated that the non-linear relation between UV and X--ray luminosity in quasars can be used as a cosmological probe. 
We have shown that within the precision allowed by present data, this relation does not show any evolution with redshift and/or luminosity, and can be assumed to remain constant at all redshifts. Based on this result, we built a Hubble diagram for quasars, which extends up to $z>6$, and is in perfect agreement with the analogous Hubble diagram for supernovae in the $z\sim0.01-1.4$ range in common. The main advantage of this method is clearly the possibility of testing the cosmological model, and measuring the cosmological parameters, over a wider redshift range than any other cosmological probe, with the possible exception of gamma-ray bursts \citep{ghirlanda2006}. The main limitation lies in the large observed dispersion of the relation ($\sigma$$\sim$0.3 in a $\log \luv$-$\log \lx$ plane). As a consequence, large samples are needed to obtain significant constraints on cosmological parameters.

With present data, we were able to build a sample of 808 quasars by merging several literature samples accurately cleaned from BAL, radio loud, and optically heavily reddened objects. A correction for dust extinction was applied to the moderately reddened ones. This sample allows a first application as a cosmological probe: assuming the $\Lambda$CDM model, we obtained $\Omega_M$=0.21$^{+0.08}_{-0.10}$, and a looser constraint on $\Omega_\Lambda$ (a lower limit of $\sim 0.7$, due to small amount of low-$z$ quasars in our sample). The constraints on the cosmological parameters are obtained by fitting a Hubble Diagram, analogously to what is done with supernovae. Therefore, we performed joint fits of our data and the Union~2.1 SNe sample of the Supernovae Cosmology Project, and obtained much tighter constraints: $\Omega_M$=0.21$^{+0.04}_{-0.04}$ and $\Omega_\Lambda$=0.74$^{+0.08}_{-0.08}$. These uncertainties are still large compared to the estimates obtained by combining all the available cosmological probes (CMB, SNe, BAO, lensing). However, they are better by about a factor of two that those from SNe alone, showing that the quasar sample available today can already provide a significant contribution to cosmological studies that are based on distance measurements only.

Finally, we discussed the potential of quasars as cosmological probes considering future samples with both UV and X--ray measurements, which will become available in the next few years. With currently available observatories, new dedicated observations of well-selected high-$z$ quasars will greatly improve the test of the cosmological model at $z>4$. The forthcoming eROSITA all-sky X--ray survey will provide X--ray measurements for more than 100,000 SDSS quasars. Further in the future, surveys from {\it Euclid} and LSST in the optical-UV, and {\it Athena} and other possible wide field X--ray survey telescope, will provide samples of millions of quasars. With these samples it will be possible to obtain constraints on possible deviations from the standard cosmological model, which will rival and complement those available from the other methods.

In future papers we will further investigate the potential of the new method presented here. We will further discuss possible limitations and systematic effects, which are likely not relevant with relatively small samples such as the one available at present, but may become significant when the increased size of the samples will allow more precise measurements.


\section*{Acknowledgments}
We thank the referee, Michael Strauss, for very helpful and constructive comments, which helped to improve the paper.
We would also like to deeply thank a number of colleagues for carefully reading our paper and provide useful comments and suggestions: Marco Salvati, Filippo Mannucci, Gianni Zamorani, Alessandro Marconi, Martin Elvis, Andrea Comastri, Roberto Gilli, and Marcella Brusa. E. L. thanks Matteo Martinelli for his help on using the Planck 2015 cosmological chains. 
E.L. thanks Vincenzo Mainieri and Marcella Brusa for comments and clarifications on the use of the results from X--ray spectroscopy in the XMM-COSMOS. 
This research made use of matplotlib, a Python library for publication quality graphics \citep{2007CSE.....9...90H}.
This work has been supported by the grant PRIN-INAF 2012.

\appendix
\section{appendix A \\Literature sample}
\label{appendixA}
We have considered the samples presented by \citet{steffen06}, \citet{shemmer06}, \citet{just07}, \citet{young10}, and \citet{lusso2010} with both optical and X--ray luminosities at 2500~\AA\ and 2~keV, respectively. 
Given that uncertainties for both optical and X--ray luminosities were not published in most of these works, we retrieved, where possible, all multi wavelength information and re-compute luminosities with their uncertainties. 
For all catalog correlations we have used the Virtual Observatory software \textsc{TOPCAT} \citep{2005ASPC..347...29T} available online\footnote{http://www.star.bris.ac.uk/$\sim$mbt/topcat/}. 
The total quasar sample considered consists on 1,138 objects. A summary of the total quasar sample is given in Table~\ref{samplecosmotot}. 

\subsection{A.1. Optically selected samples}
\label{A.1}
The Steffen et al. sample (333 sources) contains 155 objects from the SDSS-DR2 quasar catalog \citep{2004AJ....128..502A,strateva05}. For these sources we have updated the optical values using the more recent quasar catalog published by \citet{2013yCat..22060004K}. 
We obtained 133 matches (excluding BAL and radio-loud quasars) using a matching radius of 3 arcsec.  
These 133 quasars have been then cross correlated with the the ROSAT archive (104 matches with a radius of 22 arcsec). The X--ray monochromatic flux at 2 keV has been obtained by converting the ROSAT/PSPC count rates in the energy band 0.1--2.4 keV into unabsorbed 0.5--2 keV fluxes by using Webpimms assuming a power law spectrum with no intrinsic absorption and a photon index $\Gamma = 2.0$ modified by Galactic absorption \citep{2005A&A...440..775K}. 

We adopted the original X--ray data for the remaining sources, while we tried to update the optical luminosities whenever possible by cross-matching them with the available SDSS catalogs. We considered the SDSS-DR7 \citep{2011ApJS..194...45S}, BOSS-DR10 \citep{2014A&A...563A..54P}, and the SDSS catalogs available in \textsc{TOPCAT}. 
This SDSS sample of 104 quasars is combined with moderate-luminosity AGNs from the COMBO-17 survey (52 objects), a subsample of sources from the Bright Quasar Survey (BQS) quasar catalog (46 objects), low-redshift optically selected AGNs (24 Sy1s), and additional optically selected, $z > 4$ AGNs (54 sources). A summary of the $[0.5-2]$ keV limit for each subsample is provided by Steffen et collaborators in their Table~3.
The new Steffen et al. sample is so composed by 280 objects. 


We adopted a similar procedure for the quasars presented by \citet{shemmer06} and Just et al. (2007). We retrieved all multi wavelength information from infrared (WISE, UKIDSS Data Release 9, and 2MASS) to optical (SDSS-DR9) to estimate optical luminosities with their uncertainties, while we kept the original X--ray luminosities for both subsamples. 

The sample presented by Just et al. (2007) is originally composed by 34 sources. 
Their sample of highly luminous quasars has been mainly drawn from the SDSS DR3 quasar catalog \citep{schneider05}. Eleven quasars already had detections in archival X--ray data, while 21 were observed using the Advanced CCD Imaging Spectrometer (ACIS) during the Chandra Cycle 7 Guaranteed Time Observing program.  The requested snapshot exposure for each target was 4 ks, and all targets were strongly detected with $\sim10-150$ counts from 0.5 to 8 keV (see their Section~2.2 for details). Their SDSS sample thus includes 32 quasars with absolute magnitudes $M_i$ values of $-29.28$ to $-30.24$, all of which have sensitive X--ray coverage; and span the range redshift $1.5\leq z\leq4.5$. Two complementary $z>4$ quasars were added to their sample APM 08279+5255 at $z=3.91$ and HS 1603+3820 at $2.51$, leading to their ``core" sample of 34 objects. 
From this core sample we excluded 7 objects classified either BALs, RLQs, and/or lensed sources (see their Table~1). We have also neglected the object 142123.98+463317.8 at $z=3.36$ which has been classified BAL by \citet{2009ApJ...692..758G}, and the sources 170100.62+641209.0 at $z=2.74$ and 152156.48+520238.4 at $z=2.19$ which were already in the Steffen et al. sample. The final Just et al. sample considered in our analysis is thus composed by 24 objects.

\citet{shemmer06} presented Chandra observations of 21 $z > 4$ quasars mainly selected from the SDSS DR3 quasar catalog for $z<5.4$ and with near-infrared imaging and spectroscopy for higher redshifts. Nineteen quasars were targeted with Chandra during Cycles 4 and 6 with short ($3-30$ ks) X--ray observations. Three other weak (or absent) emission-line quasars (WLQs) from DR3 were included in their sample (see their Section~2 for details), but given the peculiarity of such objects we neglected them for our study. From their sample we also excluded one radio loud quasar (SDSS $J001115.23+144601.8$ at $z=4.97$), one moderate radio emitting quasar (SDSS $J144231.72+011055.2$ at $z=4.51$, also classified WLQ), and two BALs (SDSS $J104845.05+463718.3$, SDSS $J165354.61+405402.1$). The final Shemmer et al. high-redshift sample considered in our analysis is thus composed by 14 objects.

We have further increased our sample with the one published by \citet{young10}, which is composed by 327 quasars (their SPECTRA sample) selected by cross-correlating the SDSS DR5 quasar catalog with the XMM-Newton archive. They fitted three models to each XMM-Newton spectrum: (1) a single power law with no intrinsic absorption, (2) a fixed power law with intrinsic absorption left free to vary, and (3) an
intrinsically absorbed power law, with both $\Gamma$ and $\nh$ left free to vary. Any spectrum without a good fit, or with significant contribution from a strong soft excess component or absorption is excluded from the final sample. All sources have both optical and X-ray spectra with an X-ray signal-to-noise ratio $(S/N) >6$ spanning a redshift of $z=0.1-4.4$ (with an $i-$band magnitude of $15.2-20.4$, see their Section~2 for further details). We have neglected overlapping objects leading to a final sample of 278 sources. Optical luminosities have been updated for 242 objects by cross-matching them with the SDSS-DR7 quasar catalog using a matching radius of 3 arcsec. For the remaining sources (36 quasars) we considered the published optical luminosities. We also kept the X--ray luminosity values as published by Young et collaborators.

\subsection{A.2. X--ray selected sample}
\label{A.2}
We considered an updated version of the catalog already published by \citet{2010ApJ...716..348B}, which includes the photometric redshift catalog by \citet{2011ApJ...742...61S}, and new spectroscopic redshift measurements\footnote{The multi-wavelength XMM-COSMOS catalog can be retrieved from:\\ http://www.mpe.mpg.de/XMMCosmos/xmm53\_release/, version $1^{\rm st}$ November 2011.}.
We have selected 1375 X--ray sources detected in the 0.5--2~keV band at a flux larger than $5\times10^{-16}\fluxunits$ over the COSMOS area (2 deg$^2$) and for which a reliable optical counterpart can be associated (\citealt{2010ApJ...716..348B}). 
From this sample, 426 objects are spectroscopically classified as broad-line AGN on the basis of broad emission lines ($FWHM > 2000 \,{\rm km \; s^{-1}}$, ``spectro-$z$" sample hereafter) in their optical spectra. 
In order to extend our XMM-COSMOS sample to fainter magnitudes, we added to the spectro-$z$ sample a sample of 116 Type-1 AGN defined as such via SED-fitting (``photo-$z$" sample hereafter). 
The photo-$z$ sample has been selected following the same approach as in \citet[see their Section 2 for details]{2013ApJ...777...86L}. We selected all sources with a best-fit photometric classification consistent with an AGN-dominated SED (i.e., $19\leq{\rm SED-Type}\leq30$ as presented by \citealt{salvato09}). 
The X--ray luminosity values have been computed following the same approach as outlined in Section~2.1 in Lusso et al. (2013). 
Briefly, count rates in the 0.5--2 keV and 2--10 keV are converted into monochromatic X--ray fluxes in the observed frame at the geometric mean of the soft (1 keV) and hard (4.5 keV) energy bands using a Galactic column density $\nh = 2.5 \times 10^{20}$ cm$^{-2}$ (see \citealt{cappelluti09}), and assuming a photon index $\Gamma=2$ and $\Gamma=1.7$, for the soft and hard band, respectively. Fluxes are then blueshifted to the rest-frame. The rest-frame monochromatic flux at 2 keV is finally obtained by interpolation of these fluxes if the source redshift is lower than $\sim$1, by extrapolation considering the slope between the fluxes described above for higher redshifts. 
The XMM-COSMOS sample comprises 542 AGN (426 with spectro-$z$ and 116 with photo-$z$) spanning a redshift range of $z=0.041-4.255$. 


\begin{table*}
\centering
\caption{Summary of the total quasar sample. \label{samplecosmotot}}
\begin{tabular}{cccccl}
\hline\hline\noalign{\smallskip}
 Number & $[0.5-2]$ keV limit & Area & $z$ & $i$ mag & Reference  \\
              &  erg sec$^{-1}$ cm$^{-2}$ & deg$^2$  &  &  & \\
\noalign{\smallskip}\hline\noalign{\smallskip}
  280         & ...$^{\mathrm{a}}$  & ...  & $0.009-6.280$ & $12.20-23.70$ & \citet{steffen06}  \\    
  24           &   ...  & ...  & $1.760-4.610$ & $15.00-20.20$ & \citet{just07} \\   
  14           &   ...  & ...  & $4.720-6.220$ & $18.34-23.78$ & \citet{shemmer06}  \\ 
  542         &  $5\times10^{-16}$  &  2    & $0.041-4.255$  & $16.86-26.04$ & \citet{lusso2010}  \\
  278         &   ... & ...  &  $0.160-4.441$ & $15.26-20.40$& \citet{young10}  \\        
\hline\noalign{\smallskip}  
1,138       &    ...  & ...  & $0.009-6.280$ & $15.00-23.78$ & Total \\     
\hline\hline\noalign{\smallskip}               
\end{tabular}                                   
                                                
\begin{list}{}{}
\item[$^{\mathrm{a}}$]Not well defined.
\end{list}                                      
\end{table*}                                                                    

\subsection{A.3. Rest-frame optical luminosity}
\label{A.3. Rest-frame optical luminosity}

To obtain the rest-frame monochromatic luminosities at 2500\AA\ we used all the available photometry compiled in the SDSS and XMM--COSMOS catalogs. To compute the rest-frame AGN SEDs we considered the flux density (erg/cm$^2$/s/\AA) and $1$--$\sigma$ uncertainty at the effective wavelength of the filter listed in the catalogs.
Galactic reddening has been taken into account: we used the selective attenuation of the stellar continuum $k(\lambda)$ taken from \citet{1999PASP..111...63F} with $R_{\rm V}=3.1$. Galactic extinction is estimated from \citet{schlegel98} for each object. We derived total luminosities at the rest-frame frequency of the filter according to the standard formula
\begin{equation}
\nu_e L_{\nu_e} = \nu_o F_{\nu_o} 4\pi D_{\rm L}^2.
\end{equation} 
The data for the SED computation from mid-infrared to UV (upper limits are not considered) were then blueshifted to the rest-frame and no K-correction has been applied. 
We determine a ``first order" SED by using a first order polynomial function, which allows us to build densely sampled SEDs at all frequencies. This choice is motivated by the fact that a single interpolation with a high-order polynomial function could introduce spurious features in the final SED. 
In the case 2500~\AA\ is covered by no less than two data points, the $\luv$ values are extracted from the rest-frame SEDs in the $\log \nu-\log\nu L_\nu$ plane ($\nu F_\nu\propto\nu^\Gamma)$. 
If the SED is constructed by two data points (or more), but they do not cover 2500 \AA, luminosities are extrapolated by considering the last (first) two photometric data points. Finally, we corrected the F$_{UV}$ estimates taking into account the redshift-dependent contribution of emission lines to the photometric points, as described in Appendix A.4.
Uncertainties on monochromatic luminosities ($L_\nu\propto \nu^{-\gamma}$) from interpolation (extrapolation) between two values $L_1$ and $L_2$ are computed as 
\begin{equation}
\delta L = \sqrt{\left( \frac{\partial L}{\partial L_1}\right)^2 (\delta L_1)^2 + \left( \frac{\partial L}{\partial L_2}\right)^2 (\delta L_2)^2}.
\end{equation}

\subsection{A.4. Emission lines contamination}
\label{A.4}
\begin{figure}
\begin{center}
\includegraphics[angle=0,width=0.45\textwidth]{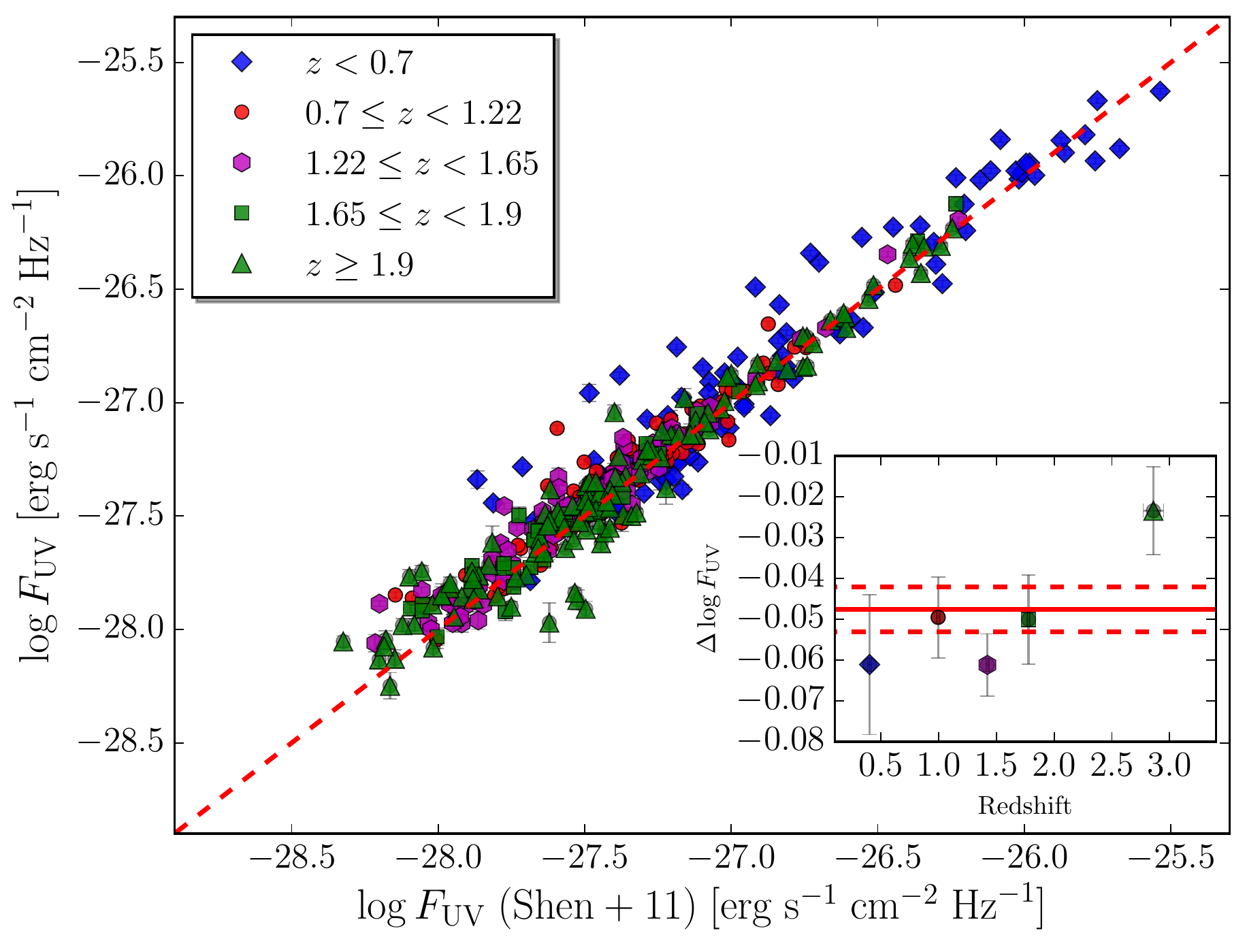}
\includegraphics[angle=0,width=0.45\textwidth]{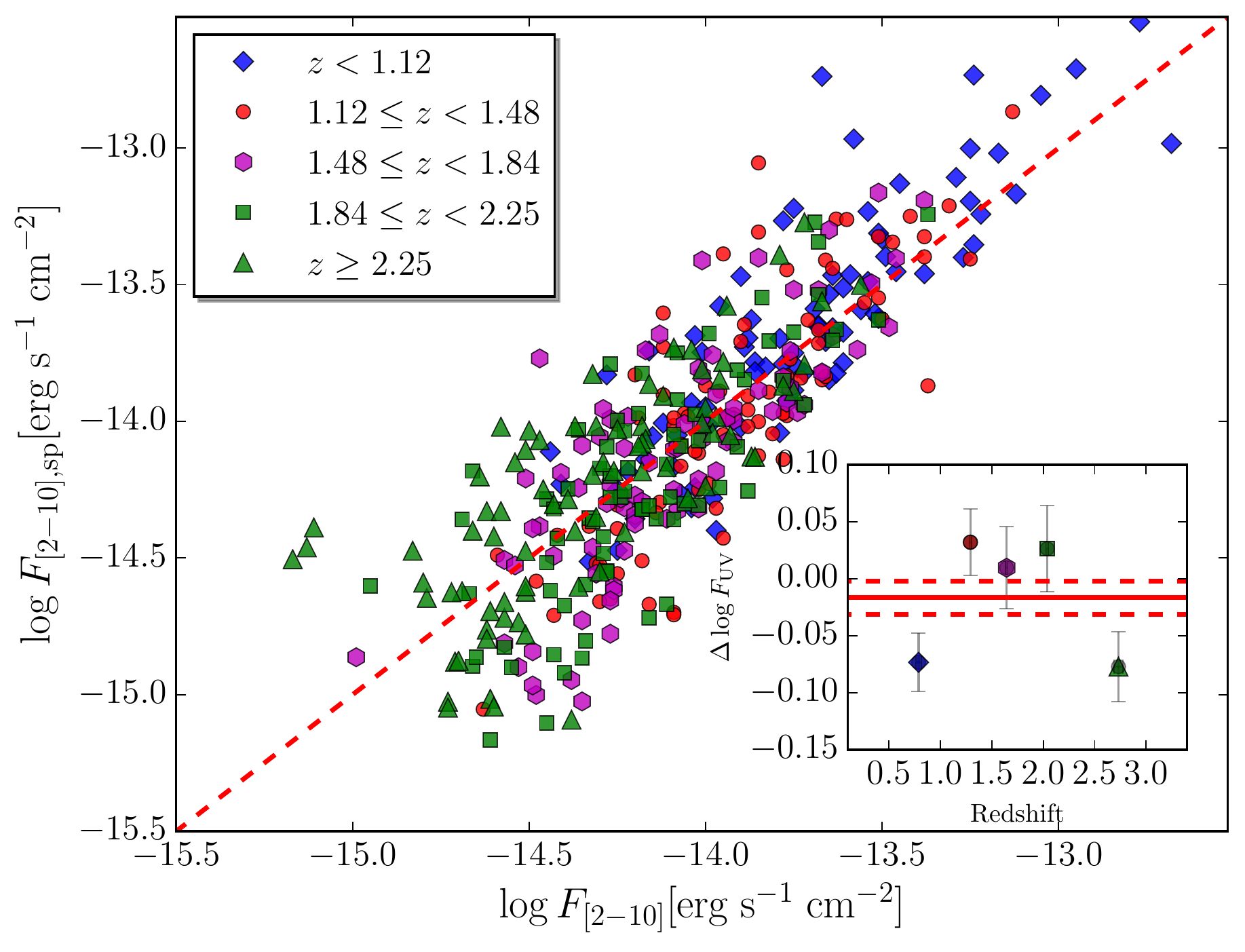}
\caption{\label{check_fo_sdss} Left panel: comparison between the optical flux measured using  described in appendix~\ref{A.3. Rest-frame optical luminosity} with the flux measured by \citet{2011ApJS..194...45S} through a complete spectral fit. Colors and point shapes represent different redshift intervals, as indicated in the plot. The dashed red line represents the one-to-one relation. The inset plot shows the average values (along with their errors) of the difference between the optical flux by \citet{2011ApJS..194...45S} and our measurements ($\Delta \log \fuv=\log F_{\rm UV, Shen+11} - \log F_{\rm UV}$) as a function of the average redshift in each bin (chosen to have approximately the same number of sources). The solid and dashed lines represent the mean and the error on the mean of the unbinned $\Delta \log \fuv$ distribution (mean is $\sim-0.05$ with $\sigma=0.12$). Right panel: same for X--ray fluxes in the 2--10 keV energy band. The comparison is between our estimates and those from the spectral analysis in Mainieri et al.~(2011). The mean over the whole sample is -0.02 with $\sigma=0.29$.} 
\end{center}
\end{figure}

Broad-band photometry of quasars usually contains a certain line contribution. 
To quantify the amplitude of this contribution we have compared the continuum flux density at rest-frame 2500 \AA\ as compiled by \citet{2011ApJS..194...45S} with our estimates obtained as described in appendix~\ref{A.3. Rest-frame optical luminosity}. Optical fluxes in \citet{2011ApJS..194...45S} were obtained through a fit of the SDSS spectra where five parameters were simultaneously fitted: the normalization and slope of the power-law continuum, and the normalization, line broadening and velocity offset relative to the systemic redshift for the iron template fit. The interested reader should refer to their Section~3 for details. We note that continuum flux measurements in the Shen et al. catalogue were neither corrected for intrinsic extinction/reddening, nor for host contamination, while our UV fluxes are corrected for dust extinction only. However, our sample selection reduces both reddening and host contaminations at minimum as discussed in Section~\ref{Source selection}. The left panel of Figure~\ref{check_fo_sdss} shows the comparison of the two $\fuv$ measures (where both are available) for 448 objects within our selected sample.  

Our flux estimates are in good agreement with the ones computed by Shen et al., yet systematically higher. 
The average values of the difference between the optical flux by \citet{2011ApJS..194...45S} and our measurements ($\Delta \log \fuv=\log F_{\rm UV, Shen+11} - \log F_{\rm UV}$) are plotted as a function of the average redshift in each bin. The solid and dashed lines represent the mean and the error on the mean of the unbinned $\Delta \log \fuv$ distribution (mean is $\sim-0.05$ with $\sigma=0.12$ in logarithm). 
A correction to the values of $\fuv$ has been applied to our data, with no significant change in the results. This effect has been also considered in the simulations, where we found that with large samples allowing more precise measurements of the cosmological parameters, the effect is not negligible (see Section 7.2).


\subsection{A.5. X-ray $K-$correction}
\label{A.5}
For 646 quasars ($\sim$57\% of the total sample) we have adopted an X--ray $K-$correction, whose systematics may affect the X--ray flux measurements and, in turn, our cosmological results. To quantify the amplitude of this possible effect, we have considered a subsample of quasars within the XMM--COSMOS sample for which the rest-frame 2--10 keV fluxes were available from a spectroscopic analysis. These fluxes are then compared with the ones computed by integrating the rest-frame X--ray SED as estimated in Section~\ref{A.2} in the 2--10 keV range.

\citet{2011A&A...535A..80M} presented a detailed spectral analysis of X--ray sources in the XMM--Newton COSMOS field. Here we consider a sample of 408 quasars with more than 70 net counts in the 0.3--10 keV energy band. The rest frame 2--10 keV fluxes ($\fxsp$) are estimated from a fit of the XMM-Newton spectra using an intrinsically absorbed power law with both $\Gamma$ and $\nh$ free to vary. The power law is then extrapolated to lower energies considering the best fit photon index, and the $\fxsp$ is finally estimated correcting the values for intrinsic absorption. 


The right panel of Figure~\ref{check_fo_sdss} shows the comparison between the rest-frame 2--10 keV fluxes from the spectral analysis done by \citet{2011A&A...535A..80M} as a function of our 2--10 keV flux measurements. We considered this correction in both our data and in the simulations. In the present sample, the effect is totally negligible. Regarding the simulated samples, the deviations found here may in principle alter the measurement of the cosmological parameters. However, the future X-ray surveys will allow a direct spectral analysis for all sources, so such systematics in the estimates of $\fx$ will be absent.


\section{appendix B\\Tests of the method}
\label{appendixB}
We tested the method used to derive the cosmological parameters, based on the Hubble diagram of quasars, with simulated quasar samples and different values of $\Omega_M$ and $\Omega_\Lambda$. We started from the $\fuv-z$ distribution of our sample of 808 quasars, and performed the following simulations: (1) we simulated $\fx$ from $(z,\fuv)$ of each quasar in our sample, assuming four different combinations of $\Omega_M$ and $\Omega_\Lambda$: ($\Omega_M$,$\Omega_\Lambda$)=(0.2,0.2), (0.2,0.8), (0.8,0.8), (0.8,0.2); (2) we repeated the simulations for samples with the same $(z,\fuv)$ distribution, but made of 50,000 objects. In this way, we test the ability of measuring different parameters of our actual sample, and also the correctness of the method in the limit of very large samples. In all cases we assume a dispersion of the $\log \luv$-$\log\lx$ relation $\delta$=0.30. The results are shown in Figure~9, and demonstrate the solidity of our approach.
\begin{figure}
\begin{center}
\includegraphics[angle=0,width=0.5\textwidth]{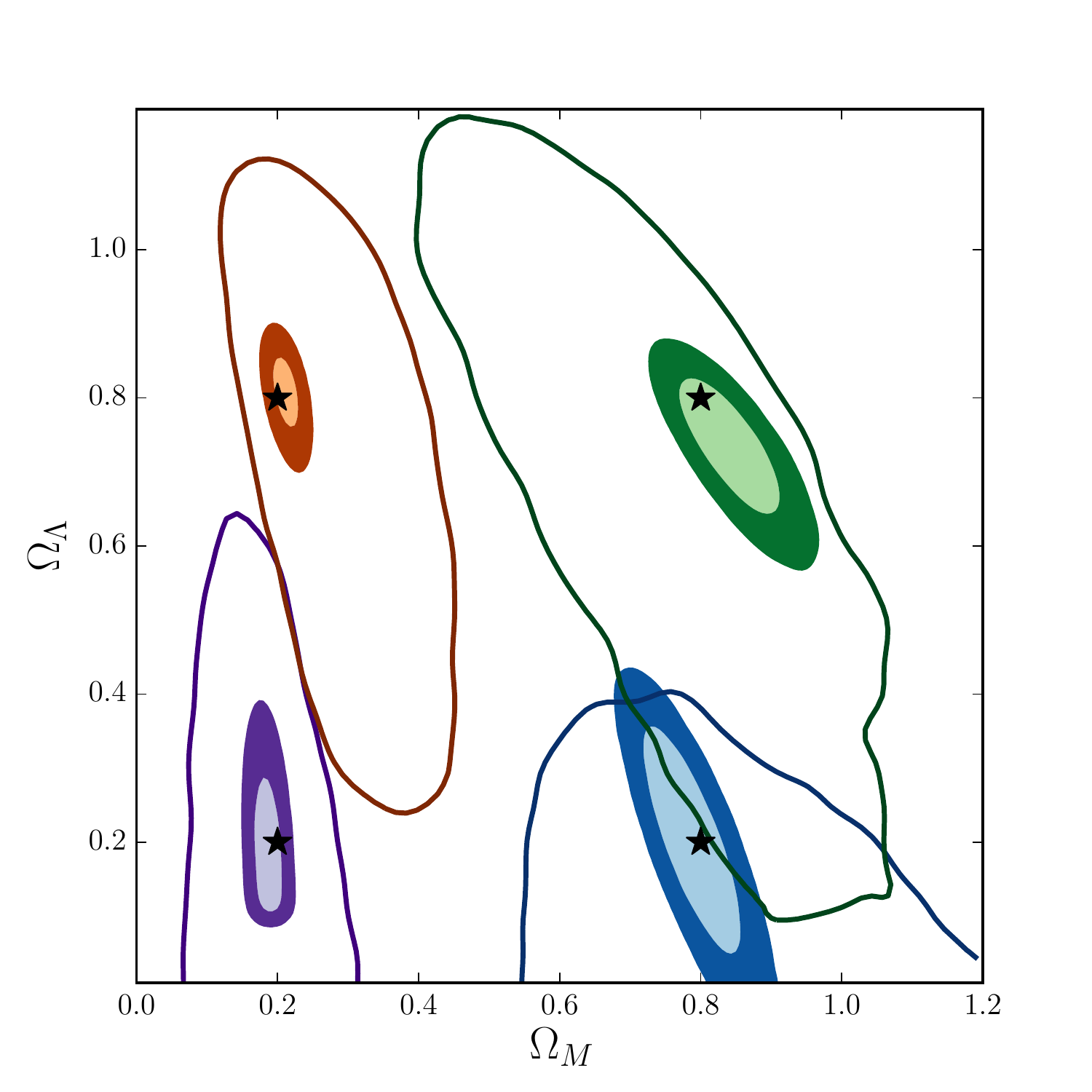}
\caption{Contours obtained fitting the Hubble diagram of quasars for simulated samples with the same size as the ``real'' one (empty contours, 68\% confidence level), and for samples of 50,000 objects (filled contours, 68\% and 95\% confidence levels), and assuming four different pairs of values for $\Omega_M$ and $\Omega_\Lambda$, as indicated by the black stars\label{app2}.}
\end{center}
\end{figure}

\bibliographystyle{apj}
\bibliography{bibl}

\end{document}